\documentclass[reprint,amssymb,amsmath,superscriptaddress,aps,showpacs,10pt,floatfix,longbibliography,prb]{revtex4-2}

\usepackage[pdftex]{graphicx}
\usepackage{epstopdf}
\usepackage{verbatim}
\usepackage{amsmath}
\usepackage{color}
\usepackage{subfigure}
\usepackage{amsbsy}
\usepackage{wasysym}
\usepackage{textcomp}
\usepackage{times}
\usepackage{float}
\usepackage{latexsym,amsmath,amssymb,bm,euscript}
\usepackage[colorlinks=true,linkcolor=black,citecolor=black,urlcolor=black]{hyperref}
\usepackage{hyperref}
\usepackage{soul}
\usepackage[normalem]{ulem}
\usepackage{mathrsfs}
\usepackage{lettrine}
\usepackage{xspace}
\usepackage{filecontents}

\begin{document}

\title{Possible coexistence of short-range resonating-valence-bond and long-range stripe correlations \\in the spatially anisotropic triangular-lattice quantum magnet Cu$_2$(OH)$_3$NO$_3$}

\author{Long Yuan}
\thanks{These authors contributed equally to this work}
\affiliation{Wuhan National High Magnetic Field Center and School of Physics, Huazhong University of Science and Technology, 430074 Wuhan, China}

\author{Yuqian Zhao}
\thanks{These authors contributed equally to this work}
\affiliation{Wuhan National High Magnetic Field Center and School of Physics, Huazhong University of Science and Technology, 430074 Wuhan, China}

\author{Boqiang Li}
\thanks{These authors contributed equally to this work}
\affiliation{Wuhan National High Magnetic Field Center and School of Physics, Huazhong University of Science and Technology, 430074 Wuhan, China}

\author{Yiru Song}
\thanks{These authors contributed equally to this work}
\affiliation{Wuhan National High Magnetic Field Center and School of Physics, Huazhong University of Science and Technology, 430074 Wuhan, China}

\author{Yuanhua Xia}
\affiliation{Key Laboratory of Neutron Physics, Institute of Nuclear Physics and Chemistry, China Academy of Engineering Physics (CAEP), Mianyang 621999, P. R. China}

\author{Benqiong Liu}
\affiliation{Key Laboratory of Neutron Physics, Institute of Nuclear Physics and Chemistry, China Academy of Engineering Physics (CAEP), Mianyang 621999, P. R. China}

\author{Junfeng Wang}
\email{jfwang@hust.edu.cn}
\affiliation{Wuhan National High Magnetic Field Center and School of Physics, Huazhong University of Science and Technology, 430074 Wuhan, China}

\author{Yuesheng Li}
\email{yuesheng\_li@hust.edu.cn}
\affiliation{Wuhan National High Magnetic Field Center and School of Physics, Huazhong University of Science and Technology, 430074 Wuhan, China}

\date{\today}

\begin{abstract}
We show that short-range resonating-valence-bond correlations and long-range order can coexist in the ground state (GS) of a frustrated spin system. Our study comprises a comprehensive investigation of the quantum magnetism on the structurally disorder-free single crystal of Cu$_2$(OH)$_3$NO$_3$, which realizes the $s$ = 1/2 Heisenberg model on a spatially anisotropic triangular lattice. Competing exchange interactions determined by fitting the magnetization measured up to 55 T give rise to an exotic GS wavefunction with coexistence of the dominant short-range resonating-valence-bond correlations and weak long-range stripe order (ordered moment $M_0$ = $|\langle s_i^z\rangle|$ $\sim$ 0.02). At low temperatures, a first-order spin-flop transition is visible at $\sim$ 1-3 T. As the applied field further increases, another two magnetic-field-induced quantum phase transitions are observed at $\sim$ 14-19 and $\sim$ 46-52 T, respectively. Simulations of the Heisenberg exchange model show semi-quantitative agreement with the magnetic-field modulation of these unconventional phases, as well as the absence of visible magnetic reflections in neutron diffraction, thus supporting the GS of the spin system of Cu$_2$(OH)$_3$NO$_3$ may be approximate to a quantum spin liquid. Our study establishes structurally disorder-free magnetic materials with spatially anisotropic exchange interactions as a possible arena for spin liquids.

\end{abstract}

\maketitle

\section{Introduction}

The $s$ = 1/2 triangular-lattice Heisenberg model (THM) is a prototypical example of the geometrical frustration. Back in 1973, the isotropic THM was initially believed to host a resonating-valence-bond (RVB) quantum spin liquid (QSL) ground state (GS), as proposed by Anderson~\cite{anderson1973resonating}. Specifically, the nearest-neighbor (NN) RVB state is a superposition of all the NN valence-bond (VB) states, where each spin $i$ forms a VB or singlet with one ($j$) of its nearest neighbors, ($\mid$$\uparrow_i\downarrow_j\rangle$$-$$\mid$$\downarrow_i\uparrow_j\rangle$)/$\sqrt{2}$~\cite{balents2010spin,RevModPhys.89.025003}. Since its initial proposal, the RVB/QSL state has been becoming one of the leading issues in condensed matter physics, due to its intimate relation to the problem of high-temperature superconductivity~\cite{anderson1987resonating} and its potential application in topological quantum computation~\cite{nayak2008non}.

However, recent numerical studies have consistently found that the GS of the  isotropic THM is actually long-range 120$^{\circ}$ N\'eel ordered~\cite{PhysRevLett.68.1766,goetze2016ground}.  The calculated GS sublattice magnetization (order parameter) $M_0^\mathrm{iso}$ = 0.205$\pm$0.015~\cite{PhysRevLett.82.3899,PhysRevB.74.224420,PhysRevLett.99.127004} is smaller than the classical value $s$ = 1/2, suggesting the survival of spin frustration and quantum fluctuations. Therefore, scientists still keep searching for QSL phases by adjusting the many-body interactions in the $s$ = 1/2 THM. For example, including second-neighbor interaction ($J_2$)  may result in the gapless $U$(1) Dirac QSL behavior~\cite{doi:10.7566/JPSJ.83.093707,PhysRevB.92.041105,PhysRevB.91.014426,PhysRevLett.123.207203}. QSL phase/behavior may also be induced by adding multiple-spin~\cite{PhysRevB.60.1064} or ring~\cite{PhysRevB.72.045105} exchange interaction, or even quenched randomness~\cite{doi:10.7566/JPSJ.83.034714,PhysRevLett.119.157201}. Motivated by the experimental findings of Cs$_2$CuCl$_4$, $\kappa$-(ET)$_2$Cu$_2$(CN)$_3$, etc., numerical works indicate that novel GSs, including QSLs, spiral and columnar ordered states, etc., may be stabilized by the spatially anisotropic exchange interactions in the THM~\cite{PhysRevB.74.014408,PhysRevB.84.174415,PhysRevB.89.184402,PhysRevB.102.224410}.  However, most of the existing numerical works are limited to the spatial anisotropy with only two parameters ($J$ and $J'$) , and lack the quantitative descriptions of experimental observations, e.g., the thermodynamic data measured on real magnetic materials.

Experimentally, several prominent $s$ = 1/2 triangular-lattice quantum magnets have been extensively studied~\cite{Li2019YbMgGaO4}, including $\kappa$-(ET)$_2$Cu$_2$(CN)$_3$~\cite{PhysRevLett.91.107001}, EtMe$_3$Sb[Pd(dmit)$_2$]$_2$~\cite{PhysRevB.77.104413}, YbMgGaO$_4$~\cite{li2015gapless,PhysRevLett.115.167203}, Cs$_2$CuCl$_4$~\cite{PhysRevB.73.134414}, Cs$_2$CuBr$_4$~\cite{PhysRevB.67.104431}, Ba$_3$CoSb$_2$O$_9$~\cite{PhysRevLett.108.057205,PhysRevLett.109.267206}, etc. Some of them, e.g.,  $\kappa$-(ET)$_2$Cu$_2$(CN)$_3$, EtMe$_3$Sb[Pd(dmit)$_2$]$_2$ and YbMgGaO$_4$, even exhibit no conventional magnetic freezing down to the lowest achievable temperature despite the relatively strong couplings, and thus were proposed as promising QSL candidates ~\cite{PhysRevLett.91.107001,PhysRevB.77.104413,li2015gapless}. However, further characterizations of these compounds have revealed the existence of profound randomness caused by the inherent structural disorder~\cite{PhysRevLett.118.107202,PhysRevB.82.125119,PhysRevB.88.075139}, which may account for the observed gapless QSL behaviors~\cite{doi:10.7566/JPSJ.83.034714,PhysRevLett.119.157201}. The existing experimental evidence for QSL remains circumstantial, and strongly depends on theoretical interpretations~\cite{Broholmeaay0668}. Very recently, the Ising triangular-lattice antiferromagnet NdTa$_7$O$_{19}$ was proposed by Arh \emph{et al.} as a quantum spin liquid candidate without structural disorder~\cite{Arh2022the}. Great efforts are still needed to search for QSL among more real materials on the bucket list. (\romannumeral1) Structural disorder has marginal effects, so that the quantum magnetism of the real compound can be precisely modeled. (\romannumeral2)  Conventional  long-range magnetic or spin-glass order is significantly suppressed by competing interactions, to make way for QSL correlations. (\romannumeral3) The compound doesn't have strong complex interactions, e.g., interlayer couplings and antisymmetric Dzyaloshinsky-Moriya (DM) superexchanges, which make the many-body computational costs extremely high. (\romannumeral4)  Large-size high-quality single crystals are available for experimental study.

\begin{figure}
\includegraphics[width=8.7cm,angle=0]{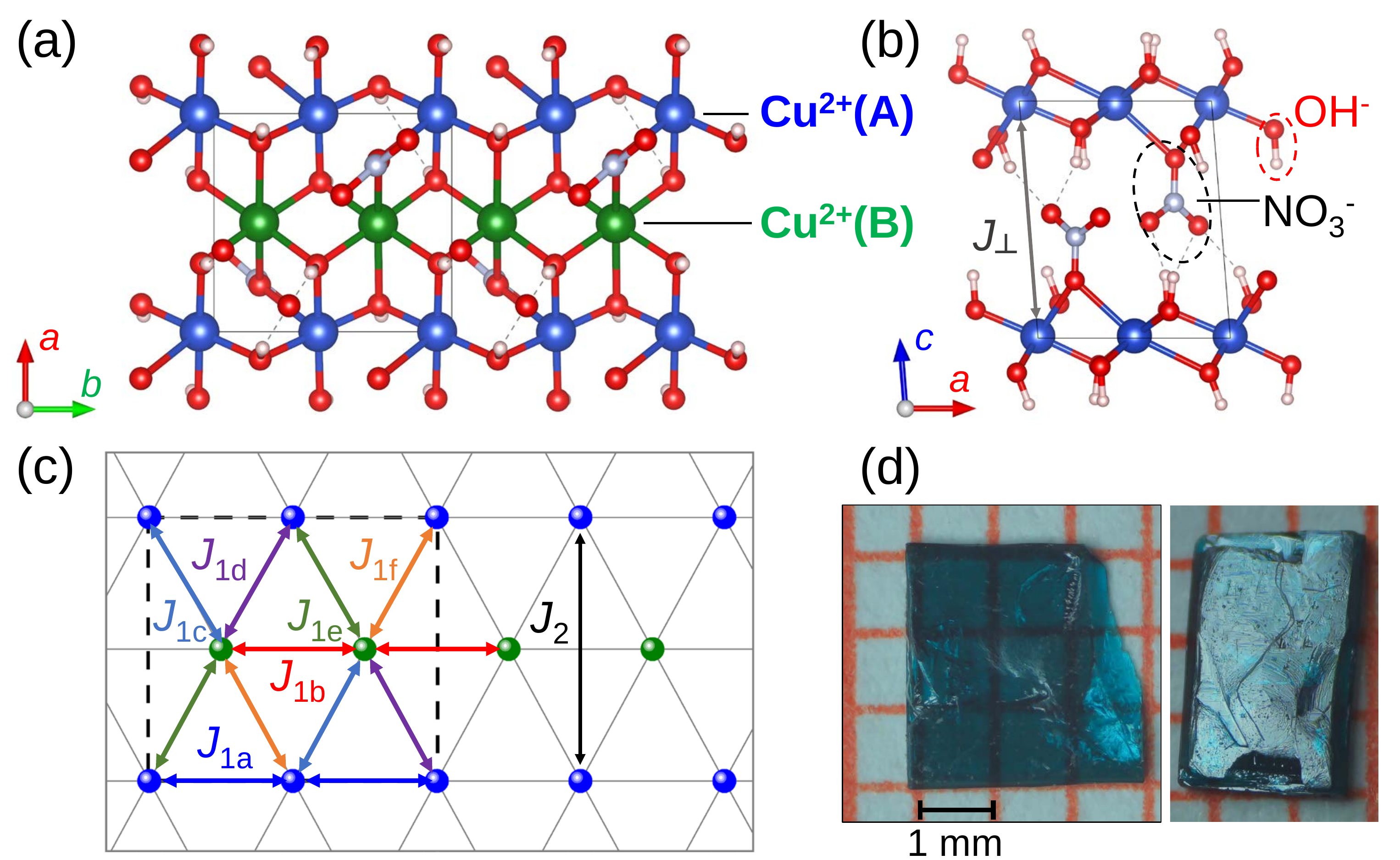}
\caption{
(a, b) Lattice structure of Cu$_2$(OH)$_3$NO$_3$ viewed along the $c$ and $b$  axes, respectively. The interlayer coupling $J_{\perp}$ is marked in (b). (c) The arrangement of the NN exchange couplings ($J_{\mathrm{1a}}$, $J_{\mathrm{1b}}$, $J_{\mathrm{1c}}$, $J_{\mathrm{1d}}$, $J_{\mathrm{1e}}$, and $J_{\mathrm{1f}}$) on the triangular lattice that is invariant under the $P$2$_1$ space group symmetry of Cu$_2$(OH)$_3$NO$_3$. The second-neighbor exchange coupling $J_{\mathrm{2}}$ is presented by the black double arrow, and the unit cell is shown by the dashed lines. (d) Typical as-grown single crystals of Cu$_2$(OH)$_3$NO$_3$.}
\label{fig1}
\end{figure}

Cu$_2$(OH)$_3$NO$_3$ (rouaite, space group $P$2$_1$) is a structurally disorder-free triangular-lattice quantum ($s$ =1/2) magnet that has been studied since a long time ago~\cite{linder1995single,drillon1995recent,ruiz2006theoretical,kikuchi2018magnetic}. The spatially anisotropic triangular-lattice spin Hamiltonian with six NN Heisenberg exchange couplings was first estimated by Ruiz \textit{et al.} using density functional theory (DFT) and quantum Monte Carlo~\cite{ruiz2006theoretical}. Recently, Kikuchi \textit{et al.} reported a long-range antiferromagnetic order at $T_\mathrm{N}$ $\sim$ 8 K according to the thermodynamic and $^1$H-NMR measurements on the powder sample~\cite{kikuchi2018magnetic}. However, such a magnetic order was not detected by powder neutron diffraction as previously reported by Drillon \textit{et al.}~\cite{drillon1995recent}. Therefore, the GS nature and quantum magnetism of the spatially anisotropic triangular-lattice compound Cu$_2$(OH)$_3$NO$_3$ remain mysterious to the community.

\begin{figure}
\begin{center}
\includegraphics[width=8cm,angle=0]{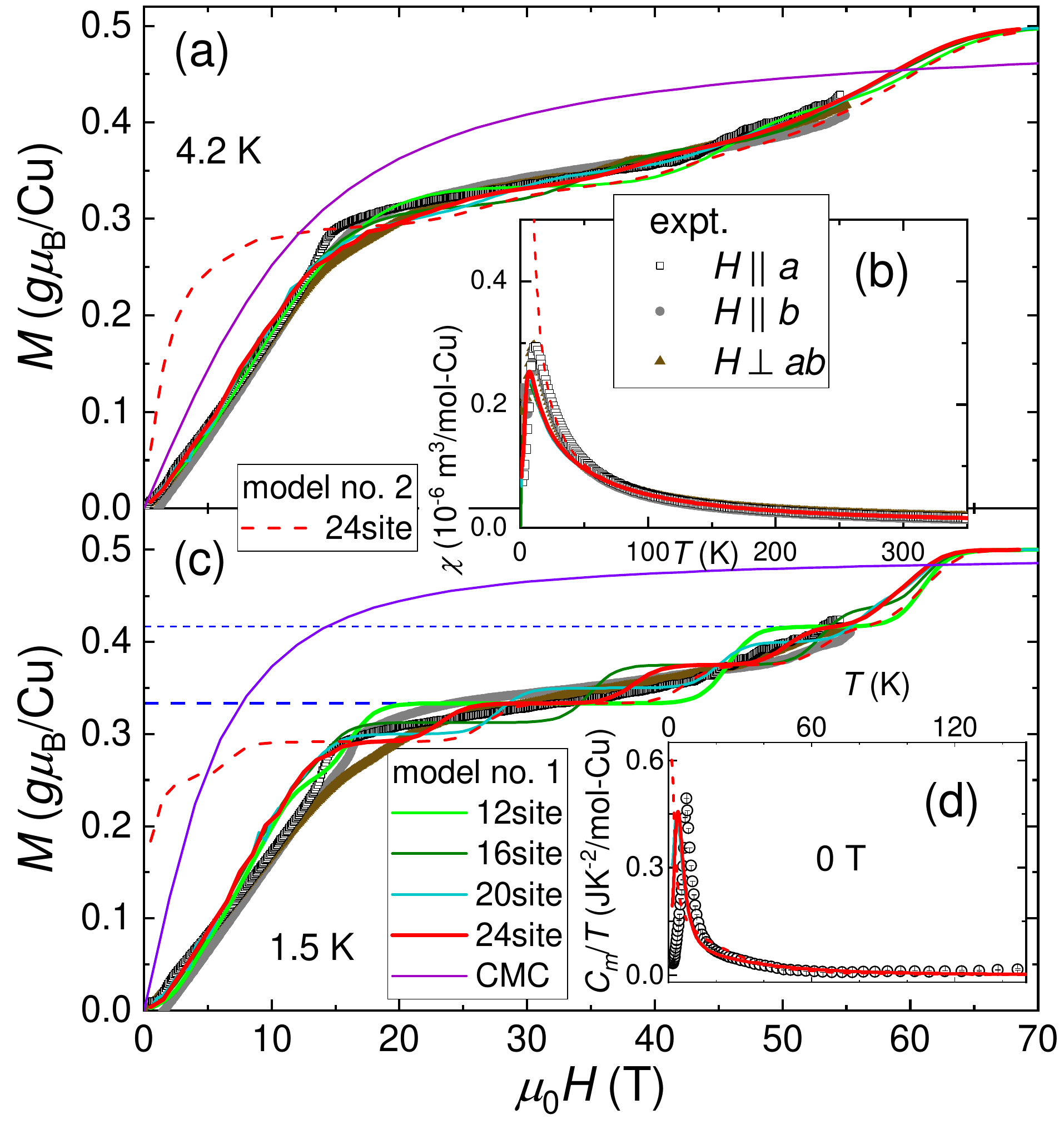}
\caption{Combined fits to (a) the isothermal field dependence of magnetization measured at 4.2 K and (b) temperature dependence of susceptibilities measured at 0.2 T, on the single crystal of Cu$_2$(OH)$_3$NO$_3$, using models 1 and 2. In model no.1, we present the quantum ($s$ = 1/2) calculations on various clusters with PBC, as well as the CMC results in the classical level. Comparison of calculations with (c) the high-field magnetization measured at 1.5 K and (d) specific heat measured at 0 T. The dashed blue lines show $\langle s_i^z\rangle$ = 2$s$/3 and 5$s$/6.}
\label{fig2}
\end{center}
\end{figure}

In this paper, we report a comprehensive investigation of the quantum magnetism on the high-quality single crystal of Cu$_2$(OH)$_3$NO$_3$, including high-field magnetization measured in pulsed fields, specific heat, dc and ac susceptibilities measured in stable fields, neutron diffraction, first-principles calculation, and quantum many-body simulation. The frustrated spin Hamiltonian with six NN exchange parameters determined by the high-field magnetization shows semi-quantitative agreement with the low-$T$ experimental observations, including the broad peaks seen in zero-field specific heat and susceptibilities, the absence of visible magnetic reflections in neutron diffraction, as well as the quantum phase transitions driven by magnetic fields at $\sim$ 1-3, $\sim$ 14-19, and $\sim$ 46-52 T, respectively. Based on the refined microscopic model, we find the zero-field GS wavefunction is a superposition of many different states in the subspace $\sum_is_i^z$ = 0 and in favor of the long-range zigzag-stripe state with a probability of $M_0$/$s$ $\sim$ 0.04. Furthermore, the GS can be almost completely represented in the NN VB basis, thus suggesting the formation of dominant short-range RVB correlations in the spin system.  Our results show that the magnetic GS of Cu$_2$(OH)$_3$NO$_3$ may be in the vicinity of a quantum spin liquid.

\section{Experimental Details}

Single crystals of $\sim$ 60\% deuterated Cu$_2$(OH)$_3$NO$_3$ [more precisely, Cu$_2$(OH$_{0.4}$D$_{0.6}$)$_3$NO$_3$] were grown by the traditional hydrothermal method. The starting materials of Cu(NO$_3$)$_2$$\cdot$3H$_2$O (10 mmol, 2.416 g), Lu(NO$_3$)$_3$$\cdot$6H$_2$O (21 mmol, 9.849 g), and 4 ml D$_2$O were mixed together and charged into a 100 ml Teflon-lined autoclave. The autoclave was heated to 503 K in a box furnace, and the temperature was maintained for three days. Transparent blue crystals with shining surfaces and a typical size of 3$\times$2$\times$0.5 mm$^3$ were obtained [see Fig.~\ref{fig1}(d)]. To see the possible sample dependence and deuteration effect, we also synthesized the non-deuterated crystals of Cu$_2$(OH)$_3$NO$_3$ in the similar way by using H$_2$O instead of D$_2$O [see Fig.~\ref{figrr1}(a)]. The good crystallization quality was confirmed by x-ray diffraction (XRD), and a set of very narrow (2$\Delta$$\theta$ $\sim$ 0.05$^{\circ}$) reflections of (0 0 $l$)  with $l$ = 1-7 were observed [Fig.~\ref{figs1}(a)]. Back-scattering Laue XRD measurements (LAUESYS\_V\_674, Photonic Science \& Engineering Ltd) were carried out to determine the orientation of the single crystal [Fig.~\ref{figs1}(c)]. The crystal structure of Cu$_2$(OH)$_3$NO$_3$ was determined by single-crystal XRD  (Mo $K_{\alpha}$, $\lambda$ = 0.71073 {\AA}, XtaLAB mini \uppercase\expandafter{\romannumeral2}, Rigaku Corporation), and the refined crystal structure was shown in Table~\ref{tabs1}.

\begin{table}
\caption{The strengths of the exchange interactions (in K) for Cu$_2$(OH)$_3$NO$_3$.}
\begin{center}
\begin{tabular}{ c || c | c | c }
\hline
\hline
& DFT+$U$ & \textbf{model no.1} & model no.2~\cite{ruiz2006theoretical} \\
\hline
$J_{1\mathrm{a}}$ & 3 & \textbf{7.2} & $-$20 \\
$J_{1\mathrm{b}}$ & 60 & \textbf{69} & 50 \\
$J_{1\mathrm{c}}$ & 7 & \textbf{1.4} & $-$9.9 \\
$J_{1\mathrm{d}}$ & $-$7 & \textbf{$-$82} & $-$15 \\
$J_{1\mathrm{e}}$ & 3 & \textbf{$-$30} & 7.1 \\
$J_{1\mathrm{f}}$ & 10 & \textbf{3.7} & $-$2.9 \\
\hline
$J_{2}$ & 0.7 & 0 & 0 \\
$J_{\bot}$ & $-$0.07 & 0 & 0 \\
\hline
\hline
\end{tabular}
\end{center}
\label{tab1}
\end{table}

Dc and ac susceptibilities of Cu$_2$(OH)$_3$NO$_3$ (up to 7 T, down to 1.8 K) were measured in a magnetic property measurement system (MPMS, Quantum Design) using well-aligned single-crystal samples of $\sim$ 70 mg. The specific heat measurements (up to 9 T, down to 1.8 K) were conducted in a Physical Properties Measurement System (PPMS, Quantum Design), using a single crystal of 4.41 mg with the $ab$ plane perpendicular to the applied magnetic field. The contributions of the N grease and puck were measured independently and subtracted from the data. The high-field magnetization data (up to 55 T, down to 1.5 K) were collected in pulsed fields using a standard inductive method with a couple of coaxial pickup coils at Wuhan National High Magnetic Field Center, and calibrated by the low-field data measured in a MPMS. Neutron powder diffraction measurements were performed on the high-resolution powder diffractometer XUANWU at China Mianyang Research Reactor down to the base temperature of 5 K using 3.2 g powder of Cu$_2$(OH$_{0.4}$D$_{0.6}$)$_3$NO$_3$, with a monochromatic neutron wavelength of $\lambda$ = 1.8846 \AA~generated from Ge(1 1 5) single crystal. The international system of units (SI) is used throughout this work.

Based on the crystal structure determined by single-crystal XRD, we conducted DFT+$U$ calculations~\cite{PhysRevB.105.024418} to estimate the strengths of exchange interactions for Cu$_2$(OH)$_3$NO$_3$, in the Vienna Ab initio Simulation Package (VASP)~\cite{vasp1,vasp2} (Appendix~\ref{sec2}).  The lattice specific heat of Cu$_2$(OH)$_3$NO$_3$ was calculated by VASP+Phonopy, and subtracted from the measured total specific heat (Appendix~\ref{sec4}). We carried out exact diagonalization (ED) and Lanczos diagonalization (LD) simulations for the thermodynamic properties, site magnetization $\langle s_i^z\rangle$, correlation function $\langle\bm{\mathrm{s}}_i$$\cdot$$\bm{\mathrm{s}}_j\rangle$, and GS wavefunction, on various clusters of $n_a$$\times$$n_b$$\times4$ = $n$ sites (spins) with periodic boundary conditions (PBC). Here, $n_a$ and $n_b$ are the lengths of the cluster in units of $a$ and $b$, respectively, and each unit cell contains 4 Cu$^{2+}$ spins [see Fig.~\ref{fig1}(c)]. By using the $S^z\equiv\sum_{j}s_j^z$ symmetry of the spin system, we performed ED calculations for the 12- and 16-site clusters with PBC. For the 20- and 24-site clusters, we carried out ED calculations in the subspaces of $S^z$ $\leq$ 5$s$ and $\geq$ ($n-$5)$s$, and then LD calculations with 50 Lanczos steps and 10 different randomly chosen states in the other subspaces~\cite{PhysRevB.105.024418,PhysRevB.98.094423}. For the 36-site cluster with PBC, we only conducted LD in the subspaces of $S^z$ =  ($n-$7)$s$ to ($n-$4)$s$ and ED in  the subspaces of $S^z$ $\ge$ ($n-$3)$s$. The finite-size effect of the calculated thermodynamic quantities is insignificant even at low temperatures among various clusters (Fig.~\ref{fig2}). The classical Monte Carlo (CMC) simulations of magnetization are conducted on a 2$\times$6$\times$4 cluster with PBC (Fig.~\ref{fig2}), where the spins are treated as classical vectors.

\begin{figure}
\begin{center}
\includegraphics[width=8.6cm,angle=0]{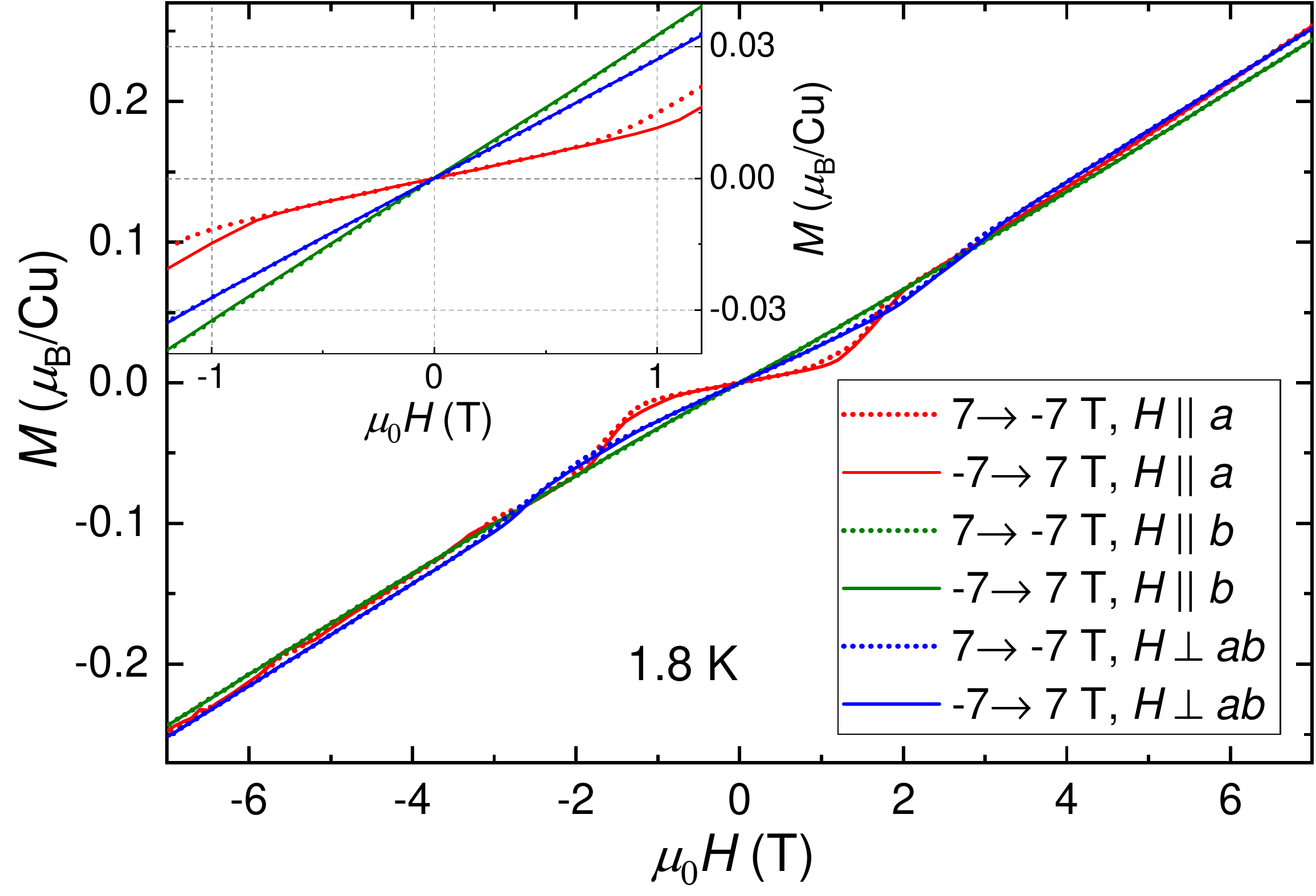}
\caption{Magnetic loops of Cu$_2$(OH)$_3$NO$_3$ measured at 1.8 K parallel to the $a$ and $b$ axes, and perpendicular to the $ab$ plane. Inset shows a zoom in plot of the data around 0 T.}
\label{figr4}
\end{center}
\end{figure}

We fit the experimental data by minimizing
\begin{equation}
R_p=\sqrt{\frac{1}{N_\mathrm{obs}}\sum_{i}(\frac{X_i^{\mathrm{obs}}-X_i^{\mathrm{cal}}}{\sigma_i^{\mathrm{obs}}})^2},
\label{eq1}
\end{equation}
where $N_\mathrm{obs}$, $X_i^{\mathrm{obs}}$, and $\sigma_i^{\mathrm{obs}}$ are the number of the data points, the observed value, and its standard deviation, respectively, whereas $X_i^{\mathrm{cal}}$ is the calculated value. In this work, we performed the combined fits to both the  field dependence of magnetization $M$($H$) and temperature dependence of susceptibilities $\chi$($T$) measured on single-crystal samples along three different directions. These experimental bases contain total 2679 data points with more than six clear features, including the zero magnetization at 0 T, slope of $M$($H$) below $H_{\mathrm{c}2}$ (see below), concrete values of the transition fields $H_{\mathrm{c}2}$ and $H_{\mathrm{c}3}$ (see below), intercept and slope of $M$($H$) at $H_{\mathrm{c}2}$ $<$ $H$ $<$ $H_{\mathrm{c}3}$, nearly zero Weiss temperature from $\chi$($T$) (see Table~\ref{tabs3}), hump temperature $T_\mathrm{p}$ $\sim$ 10 K seen in $\chi$($T$), small $\chi$($T$) at $T$ $\to$ 0 K [see Fig.~\ref{fig2}(b)],  and so on. Therefore, the experimental $M$($H$) and $\chi$($T$) should set more than six constraints in the combined fits, and all the six interaction parameters ($J_{1\mathrm{a}}$, $J_{1\mathrm{b}}$, $J_{1\mathrm{c}}$, $J_{1\mathrm{d}}$, $J_{1\mathrm{e}}$, $J_{1\mathrm{f}}$) can be well determined. In our fits, we started with three different sets of initial parameters,  the results of DFT+$U$ and previously reported model no.2 (see Table~\ref{tab1}), as well as the arbitrarily selected $J_{1\mathrm{a}}$ = $J_{1\mathrm{b}}$ = $J_{1\mathrm{c}}$ = $J_{1\mathrm{d}}$ = $J_{1\mathrm{e}}$ = $J_{1\mathrm{f}}$ = 40 K, but got nearly the same least-$R_p$ set of exchange interactions (see the model no.1 in Table~\ref{tab1}). Moreover, the experimental $M$($H$) and $\chi$($T$), as well as the aforementioned features, are successfully fitted by the minimal model no.1 with the least $R_p$ = 1.24 close to the limit $R_p$ $\sim$ 1 [see Figs.~\ref{fig2}(a) and \ref{fig2}(b)]. It is unrealistic to check $R_p$ over the full parameter region due to the computational costs of the quantum many-body calculation.

\section{Results and Discussion}

\subsection{Exchange Hamiltonian}

From simulations of the microscopic model (i.e. the exchange Hamiltonian), in principle one can understand all the observed properties of a spin system. However, the precise determination of the exchange Hamiltonian can become a key challenge in a real material, especially when there exist significant complex interactions due to the structural disorder, spin-orbit coupling, etc. The 3$d$ electron spin system of Cu$_2$(OH)$_3$NO$_3$ without evident structural disorder and strong spin-orbit coupling (see below), appears to be a perfect candidate for exploring exotic GSs.

The difference of properties among Cu$^{2+}$, OH$^{-}$, and NO$_3^-$ ions is large [see Fig.~\ref{fig1} (a)], and no site-mixing disorder was reported to date, in Cu$_2$(OH)$_3$NO$_3$. The complex magnetic effect of quenched randomness that is widespread in the existing QSL candidates~\cite{Broholmeaay0668} [e.g., the notorious site-mixing disorder of Cu$^{2+}$/Zn$^{2+}$ in ZnCu$_3$(OH)$_6$Cl$_2$] may be irrelevant to Cu$_2$(OH)$_3$NO$_3$. Both the magnetic-ion defects and nonmagnetic-ion disorder will induce quasi-free local magnetic moments, and thus give rise to the Brillouin-function-like magnetization that is a concave function of magnetic field $H$ at $H$ $\to$ 0, as well as the upturn seen in susceptibility~\cite{PhysRevB.105.024418}. In contrast, the measured magnetization of Cu$_2$(OH)$_3$NO$_3$ behaves as a convex function of $H$ below $\sim$ 1.5 T (see Fig.~\ref{fig5}), which can be well understood by our model no.1 without any quasi-free moments (see below). Moreover, the measured susceptibilities also exhibit no obvious upturn of quasi-free moments at low temperatures (see Fig.~\ref{fig4}). Therefore, in Cu$_2$(OH)$_3$NO$_3$ the structural disorder, if it exists, has marginal effects on the quantum magnetism.

\begin{figure}
\begin{center}
\includegraphics[width=8.7cm,angle=0]{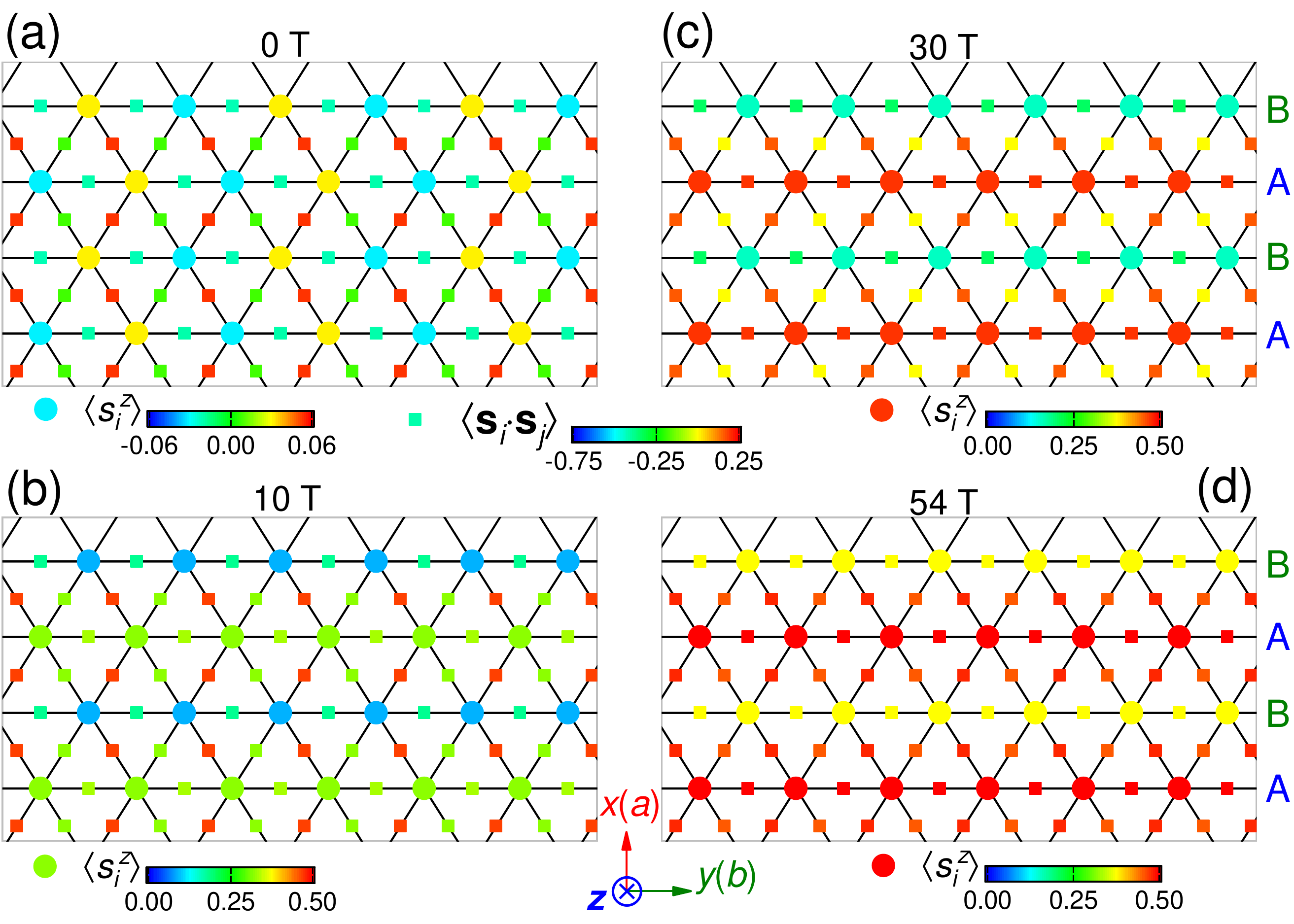}
\caption{GS magnetic structures of (a) zigzag-stripe order along the $a$ axis at 0 T, (b) LF-stripe order along the $b$ axis at 10 T, (c) possible 2/3-plateau state at $\sim$ 30 T, and (d) possible 5/6-plateau state at $\sim$ 54 T, calculated by the model no.1. The large circles present the site magnetization, whereas the small squares show the local correlation functions of the NN spin pairs sharing the same color scale. The inset defines the coordinate system for the spin components, and the A and B chains are marked.}
\label{fig3}
\end{center}
\end{figure}

Cu$_2$(OH)$_3$NO$_3$ is a good insulator with no measurable conductance at room temperature, thus the 3$d$ electron of Cu$^{2+}$ is localized spatially. The previously reported DFT work~\cite{ruiz2006theoretical} indicated that the long-distance ($\sim$ 6.9 \AA) interlayer interaction $J_{\perp}$ [see Fig.~\ref{fig1}(b)] is essentially negligible, in agreement with our DFT+$U$ result (Table~\ref{tab1}). Furthermore, our DFT+$U$ calculation suggests that the strength of second-neighbor ($\sim$ 5.3 \AA) intra-layer coupling $J_2$ is more than one order of magnitude smaller than those of the NN ($\sim$ 3.1 \AA) exchange couplings $J_1$s. Therefore, we mainly restrict ourselves to the effective model with only NN Heisenberg exchange couplings for Cu$_2$(OH)$_3$NO$_3$, as previously reported in Refs.~\cite{ruiz2006theoretical,kikuchi2018magnetic}.

The anisotropy of the exchange interaction indeed has effects on the magnetism of Cu$_2$(OH)$_3$NO$_3$ at the energy scale of $\sim$ 1 K,  because the measured magnetization shows anisotropy between $\sim$ $-$2 and 2 T (see Fig.~\ref{figr4}). However,  at higher energy scales ($|\mu_0H|$ $>$ 2 T) the anisotropy gets insignificant as shown in Fig.~\ref{figr4}. Therefore, the minimal isotropic model can provide at least a \emph{semi-quantitative} description of the quantum magnetism of Cu$_2$(OH)$_3$NO$_3$ at higher energy scales ($>$ 1 K). Of course, to account for the detailed magnetic anisotropy at the low energy scale one has to include the anisotropic exchange terms in the spin Hamiltonian. Like most Cu$^{2+}$-based oxides~\cite{PhysRevB.105.024418,PhysRevLett.101.026405}, Cu$_2$(OH)$_3$NO$_3$ contains a weak anisotropic superexchange (DM) interaction ($\sim$ 0.1$J_1$, see Appendix~\ref{sec3}), which should play a less important role. However, the inclusion of the symmetrically allowed nearest-neighbor DM terms requires 18 more parameters (each inequivalent Cu-Cu bond needs three parameters, $D_x$, $D_y$, and $D_z$). The experimental determination of this spin Hamiltonian requires at least 24 constraints, and thus seems technically unrealistic at present. Moreover, the $D_x$ and $D_y$ terms do not commute with the spin operator $S^z\equiv\sum_{j}s_j^z$ (thereby, one can no longer use the $S^z$ symmetry), and make the many-body computational costs extremely high. Therefore, including DM terms makes the problem even less tractable.

From a combined fit to the high-field magnetization measured at 4.2 K~\footnote{The finite-size effect of the calculated magnetization gets negligible above $\sim$ 4 K, as shown in Fig.~\ref{fig2}(a).} and $T$ dependence of susceptibilities [Figs.~\ref{fig2}(a) and \ref{fig2}(b)], we determine all the strengths of the NN Heisenberg couplings with the least $R_p$ = 1.24 [see Eq.~(\ref{eq1}) for the definition], as used in our model no.1 (Table~\ref{tab1}). The strengths of all six NN Heisenberg couplings first given by Ruiz \textit{et al.}~\cite{ruiz2006theoretical}, are represented as model no.2 in K (Table~\ref{tab1}). In model no.2, the couplings along the Cu$^{2+}$(A) and Cu$^{2+}$(B) chains (see Fig.~\ref{fig1}), $J_{1\mathrm{a}}$ = $-$20 K and $J_{1\mathrm{b}}$ = 50 K, are stronger than all the others. As a result, the spin system of Cu$_2$(OH)$_3$NO$_3$ was expected to be composed of the independent ferromagnetic A and antiferromagnetic B chains in the recent work reported by Kikuchi \textit{et al.}~\cite{kikuchi2018magnetic}, and the ordered moments of $\langle s^z\rangle$$_\mathrm{A}$ $\sim$ 0.5 and  $\langle s^z\rangle$$_\mathrm{B}$ $\sim$ 0 were predicted at 0 T~\cite{kikuchi2018magnetic}. Our more strict many-body computation of the full model no.2 also points to a similar GS with more precise $\langle s^z\rangle$$_\mathrm{A}$ $\sim$ 0.22 and  $\langle s^z\rangle$$_\mathrm{B}$ $\sim$ 0.00. However, such a ferromagnetic component was never observed experimentally in Cu$_2$(OH)$_3$NO$_3$. For instance, no evident remanent magnetization~\footnote{The model no.2 predicts a remanent magnetization value of 0.11$g$ $\mu_\mathrm{B}$/Cu $\sim$ 0.24 $\mu_\mathrm{B}$/Cu.} is detected on the single crystal of Cu$_2$(OH)$_3$NO$_3$ down to 1.8 K (see inset of Fig.~\ref{figr4}). Moreover, the model no.2 conflicts with the experimental magnetization below $\sim$ 15 T [Figs.~\ref{fig2}(a) and \ref{fig2}(c)], low-$T$ susceptibility [Fig. \ref{fig2}(b)] and specific heat [Fig. \ref{fig2}(d)]. In comparison, the simulations of our model no.1 show better agreement with the experimental thermodynamic data (Fig.~\ref{fig2}), as well as other observations (see below).

\begin{figure}
\begin{center}
\includegraphics[width=8.7cm,angle=0]{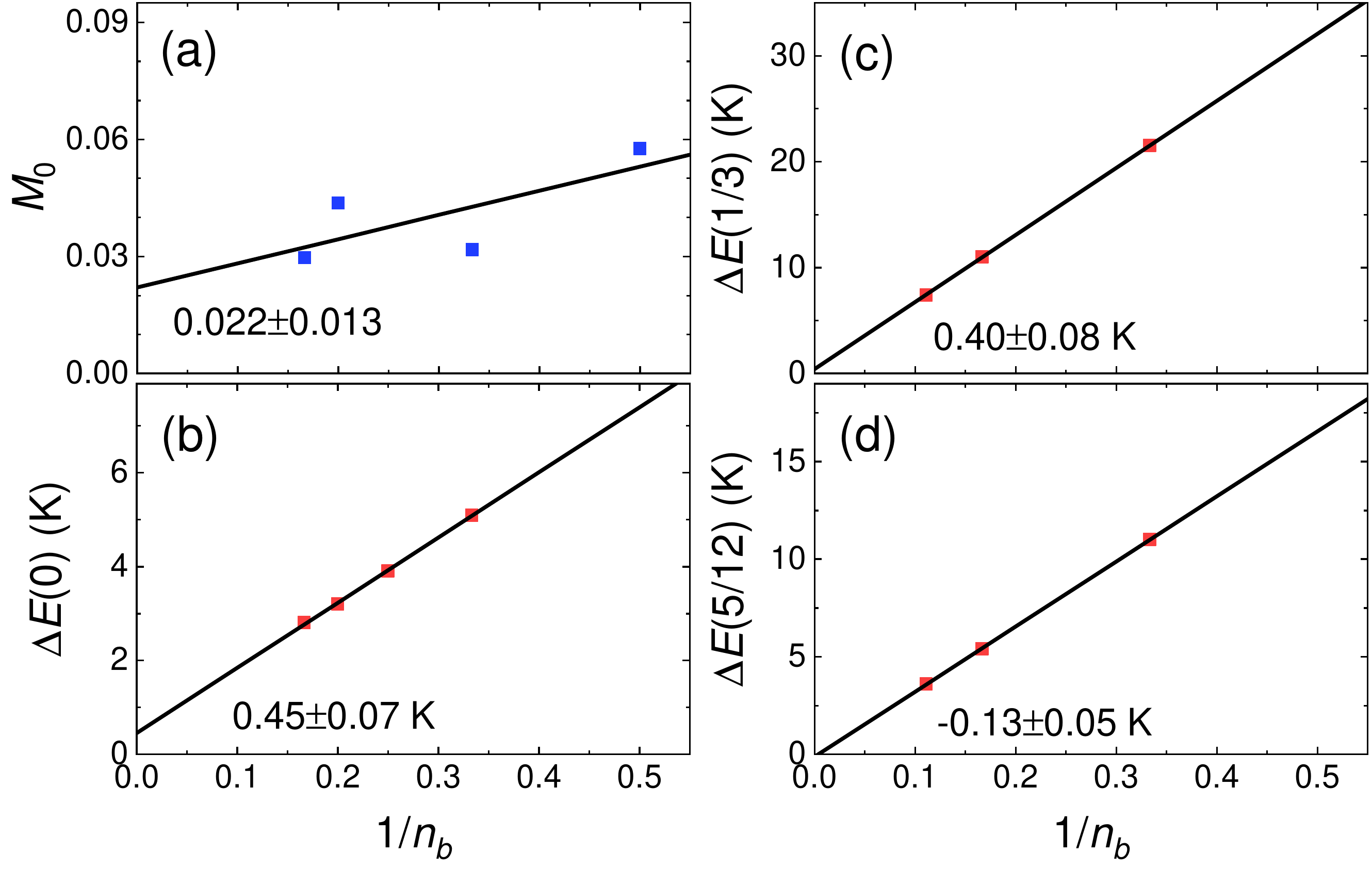}
\caption{Finite-size (1/$n_b$) dependence of (a) the zigzag stripe ordered moment $M_0$, and spin gaps at (b) $S^z$/$n$ = 0, (c) $S^z$/$n$ = 1/3, and (d) $S^z$/$n$ = 5/12, calculated by using the model no.1 along the $b$ axis. The black lines present the empirical linear fits, whose intercept values are listed. It is worth to mention that the antiferromagnetic correlations driven by $J_{1\mathrm{b}}$ are mainly along the $b$ axis.}
\label{figr1}
\end{center}
\end{figure}

\subsection{Zero-field ground-state properties}

\begin{figure}
\begin{center}
\includegraphics[width=7.5cm,angle=0]{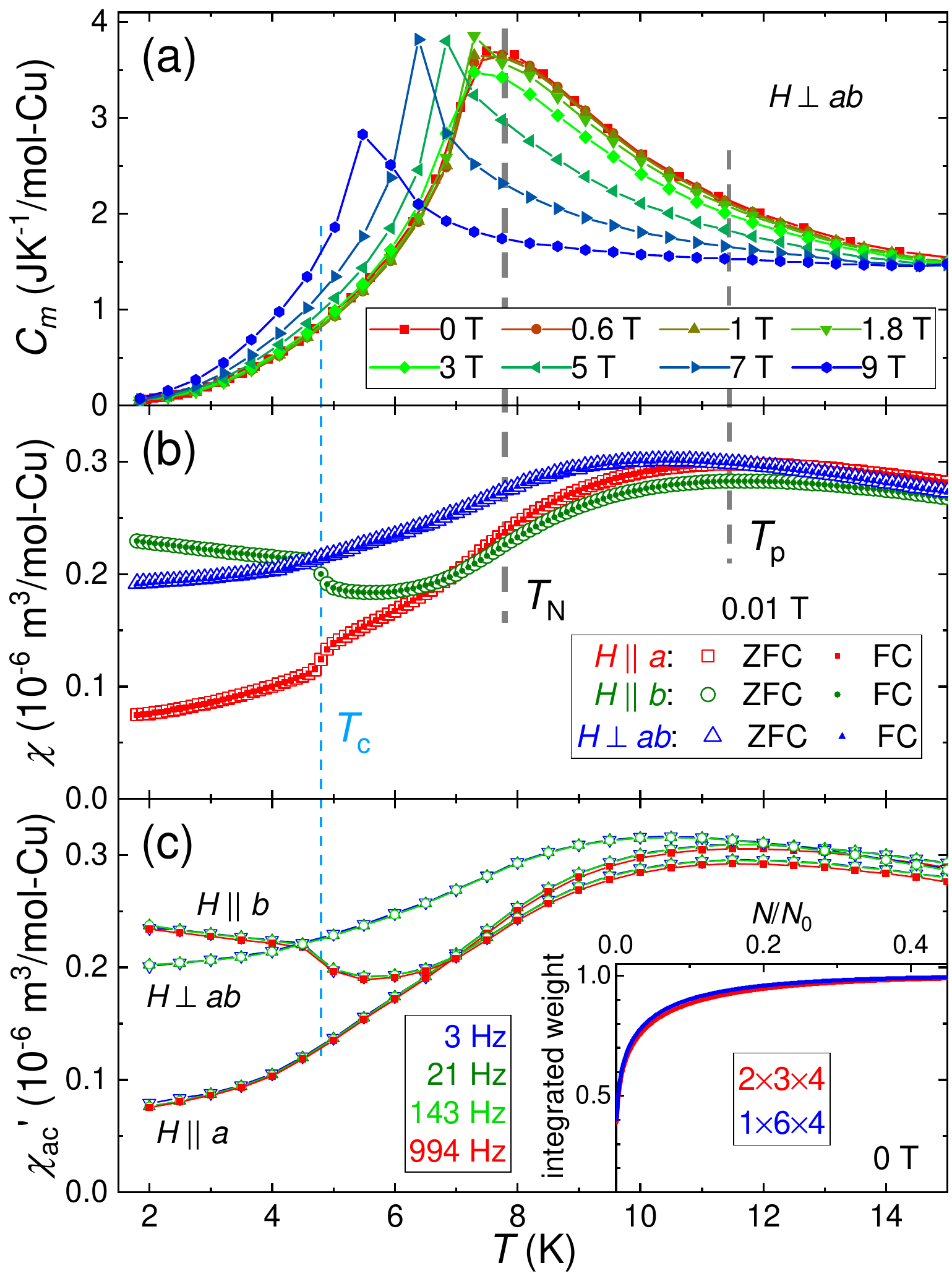}
\caption{(a) Temperature dependence of specific heat measured at various magnetic fields applied perpendicular to the $ab$ plane of Cu$_2$(OH)$_3$NO$_3$. (b) Magnetic susceptibilities measured under zero-field cooling (ZFC) and field cooling (FC) parallel to the $a$ and $b$ axes, and perpendicular to the $ab$ plane. The possible transition temperatures of $T_\mathrm{p}$, $T_\mathrm{N}$, and $T_\mathrm{c}$ are marked. (c) Temperature dependence of the ac susceptibilities (real parts) measured on the single crystal of Cu$_2$(OH)$_3$NO$_3$ down to 1.8 K. Inset of (c) shows the calculated distribution of the probabilities of most preferred states in the zero-field GS (see main text).}
\label{fig4}
\end{center}
\end{figure}

Given that model no.1 provides a semi-quantitative description of the magnetic and thermodynamic data measured on Cu$_2$(OH)$_3$NO$_3$ (see Fig.~\ref{fig2}), we explore the mysterious GS nature by diagonalizing this exchange Hamiltonian without tuning parameters. After ED, the zero-field GS wavefunction is found to be a superposition of abundant basis states in the subspace of $S^z$ $\equiv$ $\sum_is_i^z$ = 0, $\mid$GS$\rangle$ = $\sum_{k=1}^{N_0}f_k$$\mid$$k\rangle$. Here, $N_0$ = $\mathrm{C}_n^{n/2}$ is the total number of the basis states (i.e. the dimension of the subspace), and $\mid$$k\rangle$ are listed in descending order by probabilities , i.e. $|f_k|^2$ $\ge$ $|f_{k+1}|^2$. We further calculate the integrated weight $F$($N$) = $\sum_{k=1}^{N}|f_k|^2$, as a function of $N$/$N_0$ [see inset of Fig.~\ref{fig4}(c)]. Based on $F$($N$), a macroscopic number of basis states ($N$ $\ge$ 0.1$N_0$) are required to approximately ($\ge$ 90\%) produce $\mid$GS$\rangle$, at odds with the conventional (strong) long-range collinear magnetic ordering.

\begin{figure}
\begin{center}
\includegraphics[width=8.6cm,angle=0]{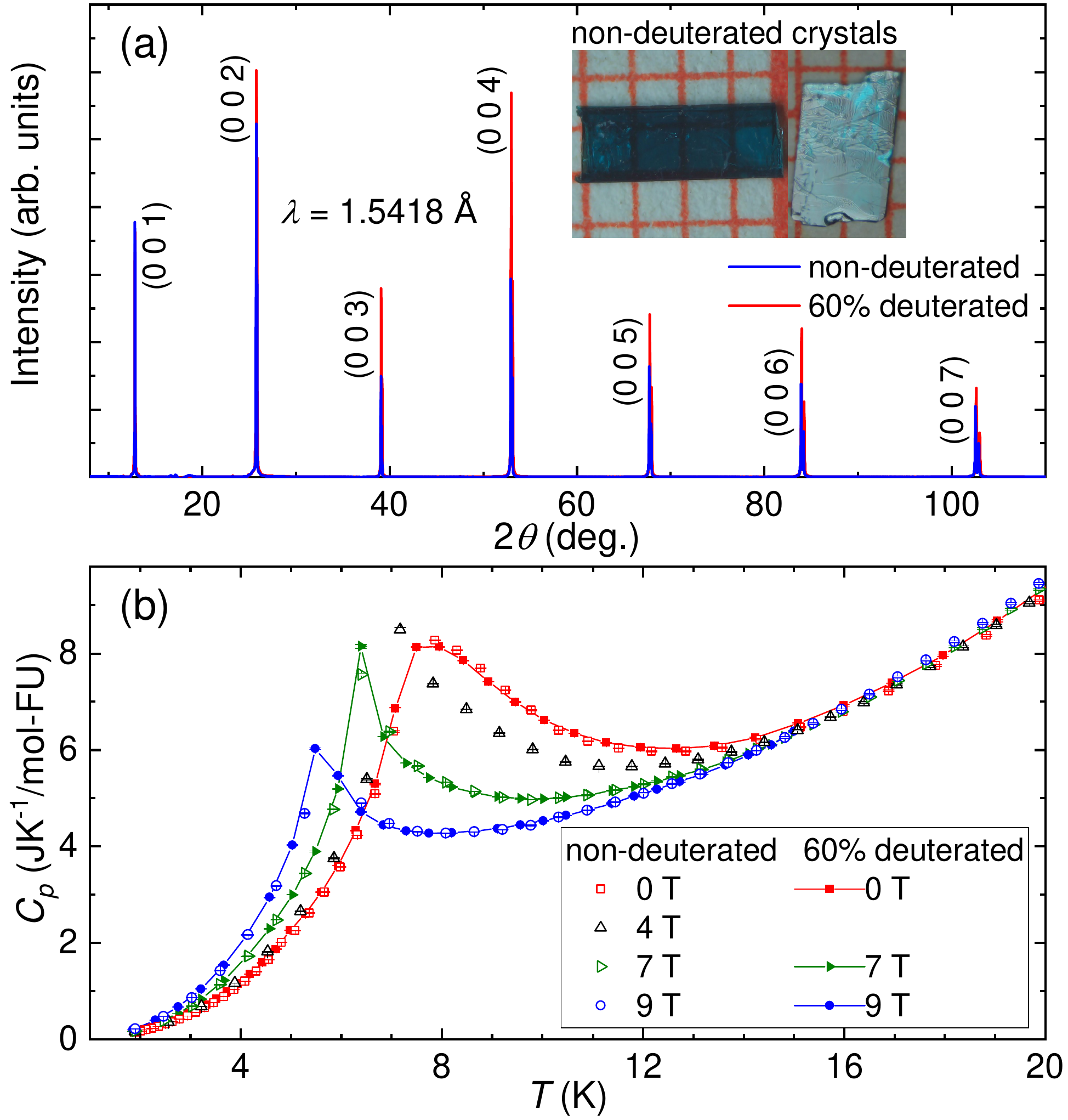}
\caption{(a) Single-crystal XRD ($\lambda$ = 1.5418 \AA) patterns measured on the $ab$-plane of the non-deuterated and 60\% deuterated crystals. The inset displays as-grown non-deuterated crystals. (b) Temperature dependence of the specific heat measured on the non-deuterated and 60\% deuterated crystals at selected fields. No significant sample dependence is detected.}
\label{figrr1}
\end{center}
\end{figure}

\begin{figure}
\begin{center}
\includegraphics[width=8.6cm,angle=0]{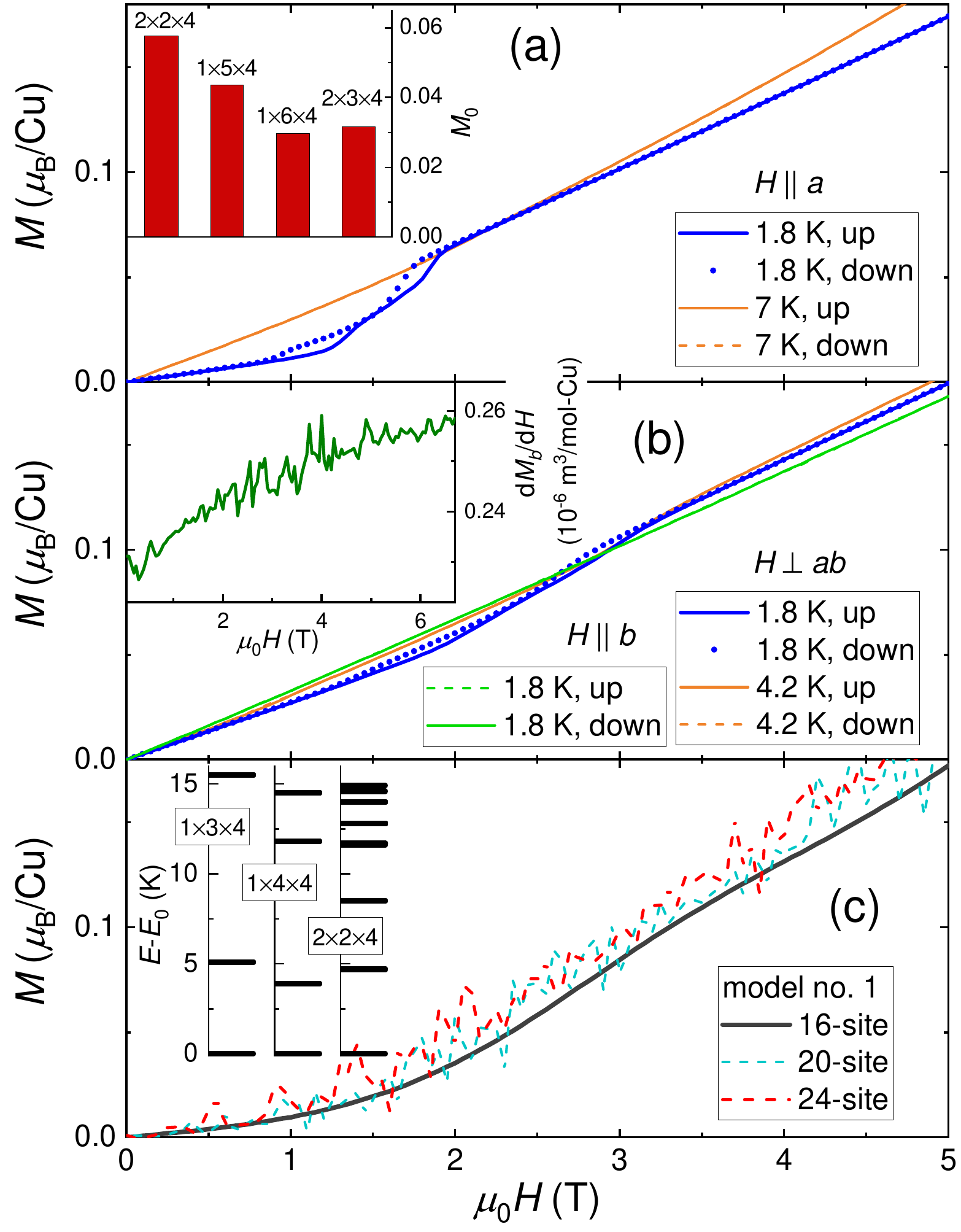}
\caption{(a, b) Magnetization of Cu$_2$(OH)$_3$NO$_3$ measured at selected temperatures parallel to the $a$ and $b$ axes, and perpendicular to the $ab$ plane, versus increasing (up) and decreasing (down) the magnetic field. (c) The field dependence of magnetization calculated at 1.8 K by the model no.1 on various clusters. The inset in (b) shows the susceptibility (d$M_b$/d$H$) measured at 1.8 K along the $b$ axis. The insets of (a) and (c) present the GS ordered moments ($M_0$ = $|\langle s_i^z\rangle|$) and low-lying energy levels, respectively, calculated on various clusters with different PBC at 0 T.}
\label{fig5}
\end{center}
\end{figure}

Among all the basis states in the $S^z$ = 0 subspace, the collinear zigzag-stripe state~\footnote{Please note that the symmetrical \{$s_i^z$\} and \{$-$$s_i^z$\} with the equal amplitude belong to the same state.} has the highest probability, $|f_1|^2$ = $M_0$/$s$ $\sim$ 0.06(2) at $n$ $\to$ 24 [see inset of Fig.~\ref{fig5}(a)], which is still tiny but more than 5 times larger than $|f_2|^2$. Therefore, the calculations of model no.1 suggest the formation of a very week long-range zigzag-stripe order with $M_0$ = 0.03(1) [see Fig.~\ref{fig3}(a)] at 0 T and low $T$, which breaks the space group symmetry ($-$$x$, $y$$+$1/2, $-$$z$) and may account for the weak antiferromagnetic order observed in Cu$_2$(OH)$_3$NO$_3$ (see below). Furthermore, the calculated NN correlation functions clearly exhibit a trend of parallel arrangement of spins along the zigzag chains with $\langle\bm{\mathrm{s}}_i$$\cdot$$\bm{\mathrm{s}}_j\rangle$ = 0.2 $\to$ 1/4 [see Fig.~\ref{fig3}(a)] because of the strongest ferromagnetic coupling $J_{1\mathrm{d}}$ [see Fig.~\ref{fig1}(c) and Table~\ref{tab1}]. This also strongly supports the formation of the long-range collinear zigzag-stripe order at 0 T and low $T$. The finite-size (1/$n_b$) dependence of the zig-zag stripe ordered moment $M_0$ suggests $M_0$ $\sim$ 0.022$\pm$0.013 in the thermodynamic limit (1/$n_b$ $\to$ 0) [see Fig.~\ref{figr1}(a)].

Our zero-field specific heat ($C_p$) measured on single-crystal samples also exhibits a broad peak at $T_\mathrm{N}$ $\sim$ 8 K [Fig.~\ref{figrr1}(b)], but its absolute value (e.g., $C_p$ $\sim$ 9.4 JK$^{-1}$/mol-FU at 20 K) is only about one third of that previously reported on powder samples ($C_p$ $\sim$ 30 JK$^{-1}$/mol at 20 K)~\cite{kikuchi2018magnetic}. To figure this out, we also carefully measured the specific heat of  the non-deuterated crystal (4.45 mg). There is no essential difference between our data independently measured on the 60\% and non deuterated crystals [see Fig.~\ref{figrr1}(b)]. Moreover, our specific heat data are consistent with  the $ab$-initio calculations at high temperatures, and the magnetic entropy estimation $\Delta S_m$($T$$\to$$\infty$) $\to$ $R$ln2 [see Fig.~\ref{figs2}(b)]. It is also worth to mention that the suspicious drop of $C_p$ between 7 and 9 T above $T_\mathrm{N}$ reported in Ref.~\cite{kikuchi2018magnetic} is clearly absent in our data [see Fig.~\ref{figrr1}(b)].

The susceptibilities also show broad humps at $T_\mathrm{p}$ $\sim$ 12 K, which is consistent with the powder result~\cite{kikuchi2018magnetic} but a bit higher than $T_\mathrm{N}$. As the temperature further decreases, another weak anomaly is observed in both ac and dc susceptibilities measured on the single crystal at $T_\mathrm{c}$ $\sim$ 5 K and $\mu_0H$ $\sim$ 0 T, which hasn't been reported in the previous powder works. These anomalies indicate the formation of antiferromagnetic order, but not a strong one, otherwise the transition temperature will be well defined and independent of probes used.  Moreover, the neutron diffraction, which is the most reliable method for characterizing long-range magnetic order so far, shows no remarkable difference between the patterns at 5 and 20 K [see Fig.~\ref{figs1}(b)], measured on the powder ground from the single crystals of Cu$_2$(OH)$_3$NO$_3$ [more precisely, Cu$_2$(OH$_{0.4}$D$_{0.6}$)$_3$NO$_3$]. The absence of any visible magnetic reflections is consistent with the result previously reported by Drillon \textit{et al.}~\cite{drillon1995recent}, and indicates that the ordered magnetic moment is small, $M_0$ $<$ 0.14~\footnote{The upper bound of the ordered magnetic moment is estimated from the measured intensity of the (0 0 2) nuclear reflection $I_\mathrm{n}^\mathrm{obs}$ = 1756 and its standard deviation $\sigma^\mathrm{obs}$ = 10, $M_0$ $<$ $\sqrt{F_\mathrm{n}^2\sigma^\mathrm{obs}/I_\mathrm{n}^\mathrm{obs}/F_\mathrm{m}^2}$, where $F_\mathrm{n}^2$ = 1.4$\times$10$^3$ fm$^2$ is the nuclear structure factor per unit cell, $F_\mathrm{m}^2$ = 408 fm$^2$ is the magnetic structure factor of the hypothetical long-range state with $s$ = 1, and 1 fm = 10$^{-15}$m~\cite{gsas}.}. Moreover, this upper bound may be overestimated, compared with the standard deviations ($\sim$ 0.01-0.04) of the ordered magnetic moments directly refined from the observed intensities of magnetic reflections (see Ref.~\cite{PhysRevB.100.144420} for example). Therefore, the GS order parameter of the zigzag-stripe antiferromagnetic state calculated by model no.1, $M_0$ $\sim$ 0.02, provides a natural explanation of the above observations. $M_0$ is even much smaller than that of the long-range 120$^{\circ}$ order stabilized in the $s$ = 1/2 isotropic THM, $M_0^\mathrm{iso}$ = 0.205$\pm$0.015~\cite{PhysRevLett.82.3899,PhysRevB.74.224420,PhysRevLett.99.127004}, thus suggesting the possible existence of much stronger spin frustration and quantum fluctuations in the anisotropic magnet Cu$_2$(OH)$_3$NO$_3$.

The magnetic entropy increase up to $T_\mathrm{N}$ $\sim$ 8 K is estimated to be only $\Delta S_m$($T_\mathrm{N}$) $\sim$ 22\%$R$ln2 [see the inset in Fig.~\ref{figs2}(b)], and it keeps increasing above $T_\mathrm{N}$, in Cu$_2$(OH)$_3$NO$_3$. Similarly, $\Delta S_m$($T_\mathrm{N}$) $\sim$ 25\%$R$ln2 and 20\%$R$ln2 were reported in the triangular-lattice antiferromagnet CeCd$_3$As$_3$~\cite{PhysRevB.103.L180406} and the honeycomb Kitaev system $\alpha$-RuCl$_3$~\cite{Do2017}, respectively. Therefore, $\Delta S_m$($T_\mathrm{N}$) of Cu$_2$(OH)$_3$NO$_3$ is comparable to those of many other two-dimensional frustrated compounds, also supporting the presence of spin frustration.

The low-energy spectra of the 1$\times$3$\times$4, 1$\times$4$\times$4, and 2$\times$2$\times$4 clusters with PBC [see inset of Fig.~\ref{fig5}(c)] were calculated by ED of model no.1, and show a qualitatively consistent result. The GS is a superposition state in the $S^z$ = 0 subspace, whereas the first excited states are three-fold degenerate in three different subspaces of $S^z$, $S^z\pm$1, respectively. For larger clusters, we have to turn to the LD  algorithm, where both the GS wavefunction and energy in each subspace can also be calculated precisely. The zero-field energy gap between the ground and first excited states (i.e. the difference between GS energies in the $S^z$ = 0 and $\pm$1 subspaces) decreases with the increase of the cluster size, but seems to have a nonzero value at $n$ $\to$ 24, $\Delta E$(0) $\sim$ 2.8 K [see Table~\ref{tab2}]. This quasi-spin-gapped behavior (triplet, $\Delta S^z$ = $\pm$1) of the GS is visible in the low-$T$ magnetization ($M$) measured on the single crystal of Cu$_2$(OH)$_3$NO$_3$, which is a convex function of $H$ at low $H$ [Fig.~\ref{fig5}, including $M_b$ measured along the $b$ axis, please see d$M_b$/d$H$ in the inset of Fig.~\ref{fig5}(b)] and semi-quantitatively described by model no.1 [Fig.~\ref{fig5}(c)]. On the other hand,  the very weak long-range antiferromagnetic order ($M_0$ $\sim$ 0.02) suggests the existence of very rare low-lying energy levels (i.e. gapless excitations with a low density of states) according to the Goldstone's theorem, and thus the nonzero values of calculated $\Delta E$(0) may be a finite-size effect. Indeed, the calculated spin gap is expected to be much smaller, $\Delta E$(0) $\sim$ 0.45$\pm$0.07 K, in the thermodynamic limit, as shown in Fig.~\ref{figr1}(b).

\begin{figure}
\begin{center}
\includegraphics[width=8.6cm,angle=0]{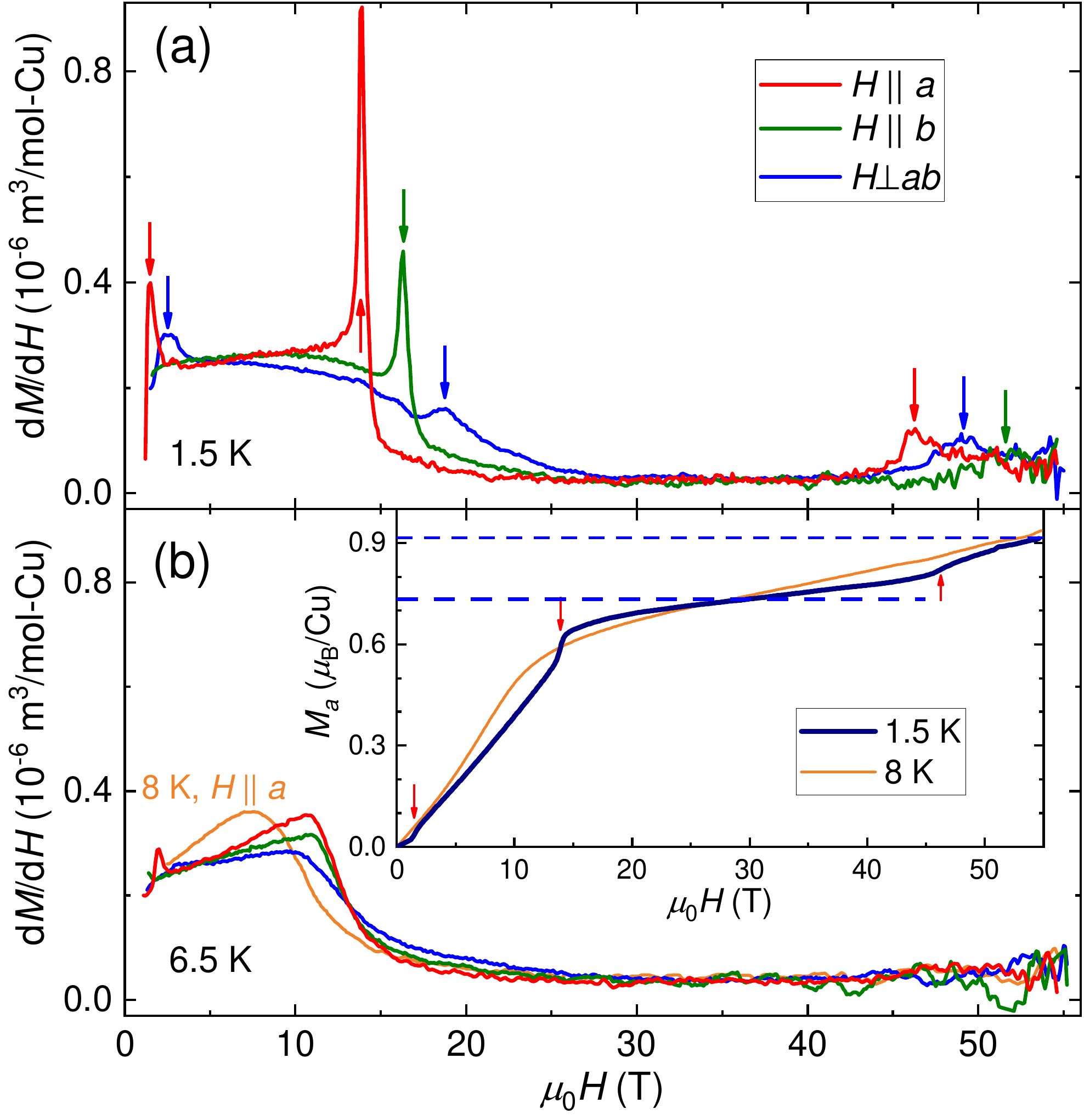}
\caption{Magnetic field dependence of susceptibilities (d$M$/d$H$) measured at (a) 1.5 K and (b) higher temperatures, on the single crystal of Cu$_2$(OH)$_3$NO$_3$. The colored arrows mark the transition fields observed at 1.5 K. The inset of (b) shows the magnetization measured along the $a$ axis, and the dashed lines display $M_a$ = $g_a$/3 and 5$g_a$/12.}
\label{fig6}
\end{center}
\end{figure}

\subsection{Field-induced magnetic transitions}

Asymmetric magnetic hysteresis is clearly observed in the low-$T$ isothermal magnetization measured with the field  applied along the $a$ axis [see Fig.~\ref{fig5}(a)], indicating there exists a first-order  magnetic transition (spin-flop transition) at $\sim$ 0.8-2 T.  Perpendicular to the $ab$ plane, the low-$T$ magnetic hysteresis gets less apparent and occurs at slightly higher fields of $\sim$ 1.2-3.4 T [see Fig.~\ref{fig5}(b)]. The spin-flop transition at $H_{\mathrm{c}1}$ becomes invisible, only when the magnetic field is applied along the $b$ axis. As the applied magnetic field increases further, another two magnetic transitions are detected at $\mu_0H_{\mathrm{c}2}$ $\sim$ 13.8-18.6 T and $\mu_0H_{\mathrm{c}3}$ $\sim$ 46.2-51.6 T, respectively, in the $H$ dependence of susceptibilities (d$M$/d$H$) measured at 1.5 K along all three different directions [see Fig.~\ref{fig6}(a)].  Above $\sim$ 6.5 K, these field induced magnetic transitions are almost undetectable in the d$M$/d$H$ curves, as shown in Fig.~\ref{fig6}(b). The spin system gets fully polarized above $\mu_0H_{\mathrm{s}}$ $\sim$ 63 T extrapolated from model no.1 [Fig.~\ref{fig2} (c)], and we focus on different GSs stabilized by external magnetic fields of $\sim$ 10, 30, and  54 T, respectively.

\begin{table}
\caption{GS energies per site ($E_0$/$n$), magnetization ($S^z$/$n$), and spin (triplet, $\Delta S^z$ = $\pm$1) gaps calculated under selected magnetic fields by using model no.1 on various clusters with different PBC.}
\begin{center}
\begin{tabular}{ c || c | c | c | c}
\hline
\hline
 & 1$\times$3$\times$4 & 2$\times$3$\times$4 & 1$\times$6$\times$4 & 1$\times$9$\times$4 \\
\hline
0 T, $E_0$/$n$ (K) & $-$21.28 & $-$21.03 & $-$21.00 & - \\
$S^z$/$n$ & 0 & 0 & 0 & - \\
$\Delta E$($S^z$) (K) & 5.1 & 2.8 & 2.8 & - \\
\hline
10 T, $E_0$/$n$ (K) & $-$22.41 & $-$22.37 & $-$22.33 & - \\
$S^z$/$n$ & 1/6 & 5/24 & 5/24 & - \\
$\Delta E$($S^z$) (K) & 0.4 & 1.1 & 1.2 & - \\
\hline
$\sim$ 30 T, $E_0$/$n$ (K) & $-$31.32 & $-$31.35 & $-$30.96 & $-$30.89 \\
$S^z$/$n$ & 1/3 & 1/3 & 1/3 & 1/3 \\
$\Delta E$($S^z$) (K) & 21.5 & 21.4 & 11.0 & 7.4\\
\hline
$\sim$ 54 T, $E_0$/$n$ (K) & $-$43.93 & $-$43.94 & $-$43.74 & $-$43.70\\
$S^z$/$n$ & 5/12 & 5/12 & 5/12 & 5/12 \\
$\Delta E$($S^z$) (K) & 11.0 & 10.8 & 5.4 & 3.6 \\
\hline
\hline
\end{tabular}
\end{center}
\label{tab2}
\end{table}

To explore the exact nature of these magnetic-field-induced phases, we also fall back on model no.1. Although the detailed orientation dependence of the magnetization clearly cannot be explained by Heisenberg exchange interactions, we believe that our minimal model no.1 nevertheless semi-quantitatively describes the overall trend of the $M$-$H$ curves (see Figs.~\ref{fig2} and \ref{fig5}). At $\sim$ 10 T, the calculated site magnetization, $\langle s_i^z\rangle$ = $\langle$GS$\mid$$s_i^z$$\mid$GS$\rangle$, shows a long-range collinear stripe ferrimagnetic order along the $b$ axis, i.e., the magnetic-ordering wave vector \textbf{Q}($H_{\mathrm{c}1}$$\le$$H$$\le$$H_{\mathrm{s}}$) is [1, 0, L], with $\langle s^z\rangle$$_\mathrm{A}$ $\sim$ 0.32 and $\langle s^z\rangle$$_\mathrm{B}$ $\sim$ 0.09 [Fig.~\ref{fig3}(b)]. In contrast, the calculated zero-field zigzag-tripe GS magnetic order is much weaker, $M_0$ $\sim$ 0.02, and along the $a$ axis with \textbf{Q}(0 T) = [1/2, 1, L] (see above). The profound change of \textbf{Q} from the zigzag-stripe phase at 0 T to the low-field (LF) stripe one at $\sim$ 10 T may account for the spin-flop transition observed at $\mu_0H_{\mathrm{c}1}$ $\sim$ 0.8-3.4 T (see Fig.~\ref{fig5}). When the applied magnetic field exceeds the critical value $H_{\mathrm{c}1}$, the zigzag-stripe antiferromagnetic spin configuration [see Fig.~\ref{fig3}(a)] is driven into the collinear ferrimagnetic order [see Fig.~\ref{fig3}(b)] at low temperatures. The model no.1 suggests the possible existence of the weak antiferromagnetic order, and semi-quantitatively describes the spin-flop transition (see Fig.~\ref{fig5}).

\begin{figure*}
\begin{center}
\includegraphics[width=15.5cm,angle=0]{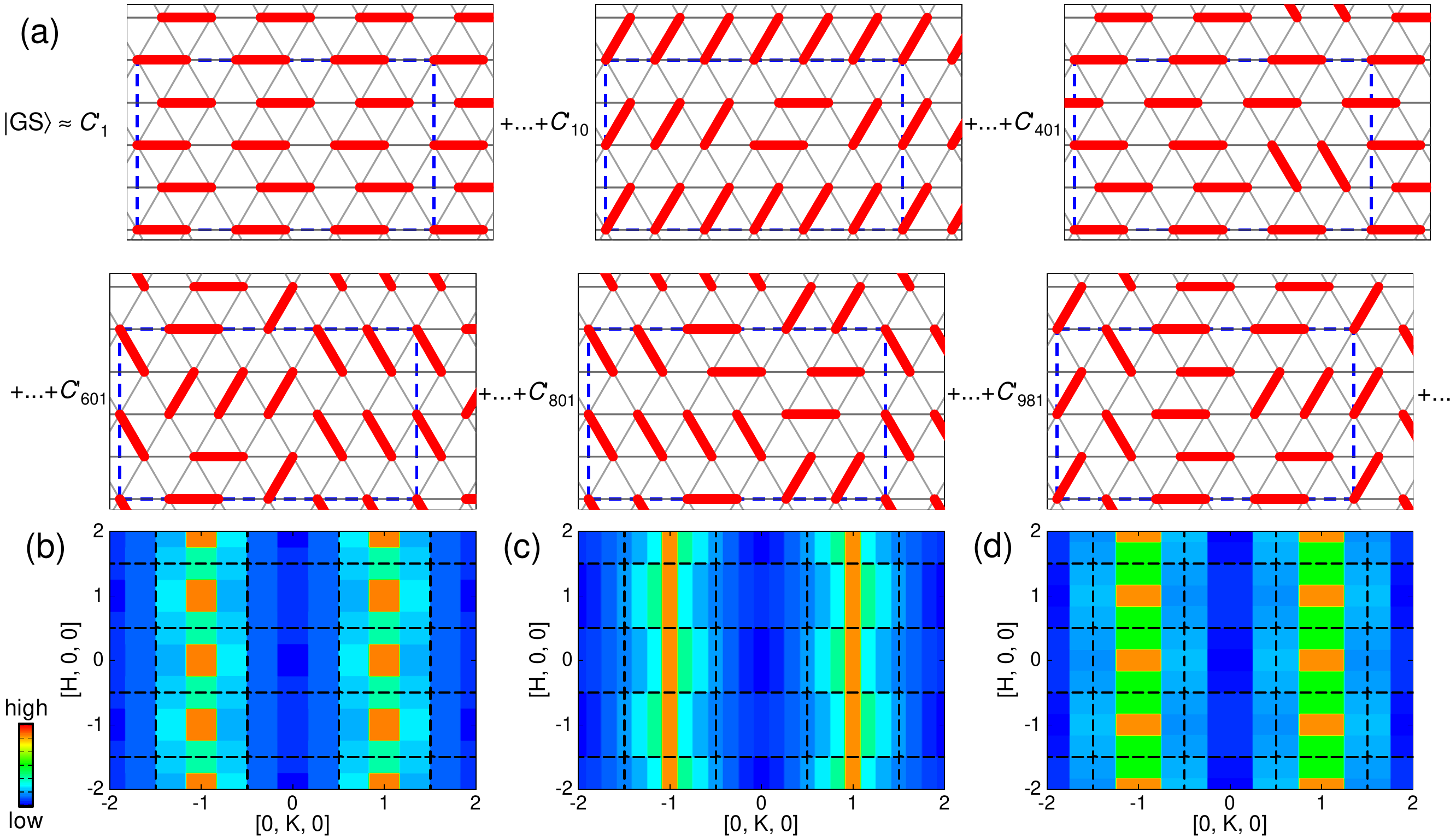}
\caption{Zero-field GS wavefunction of model no.1 approximately represented in the NN VB basis. The thick red lines mark the NN VBs or singlets, the dashed blue lines display the 2$\times$3$\times$4 cluster with PBC, and the amplitudes of the selected NN VB states are $|C'_1|$ = 4.8/$\sqrt{N_\mathrm{VB}}$, $|C'_{10}|$ = 4.2/$\sqrt{N_\mathrm{VB}}$, $|C'_{401}|$ = 3.0/$\sqrt{N_\mathrm{VB}}$, $|C'_{601}|$ = 2.9/$\sqrt{N_\mathrm{VB}}$, $|C'_{801}|$ = 2.8/$\sqrt{N_\mathrm{VB}}$, $|C'_{981}|$ = 2.7/$\sqrt{N_\mathrm{VB}}$, respectively. The wave-vector dependence of the structure factor calculated for the ground state of the model no.1 [see Eq.~(\ref{eqr1})] on (b) the 2$\times$3$\times$4, (c) 1$\times$6$\times$4, and  (d) 3$\times$2$\times$4 clusters with PBC. The dashed black lines represent Brillouin zone boundaries.}
\label{fig7}
\end{center}
\end{figure*}

The quantum simulations of model no.1 with the cluster size $n$ divisible by 12 consistently suggest the possible formation of two fractional magnetization plateaus in the subspaces of $S^z$ = $n$/3 and 5$n$/12 at $\sim$ 30 and 54 T, respectively [see Fig.~\ref{fig2}(c) and Table~\ref{tab2}]. The first magnetization plateau between $H_{\mathrm{c}2}$ and $H_{\mathrm{c}3}$ is clearly observed in the magnetization measured at 1.5 K [please see inset of Fig.~\ref{fig6}(b) for clarity], whereas the second plateau above $H_{\mathrm{c}3}$ is less visible. When the strength of the applied magnetic field approaches the maximum $\sim$ 55 T, the signal-to-noise ratio of magnetization gets relatively low, due to the slow change of the pulsed field strength and the maximum vibrating amplitude. Moreover, the field range of the second plateau seems to be very short even if it exists, according to the model no.1.

We evaluate the stabilization  of the putative fractional magnetization plateau by the spin gap defined as, $\Delta E$($S^z$)  = [$E_0$($S^z$$+$1)+$E_0$($S^z$$-$1)]/2$-$$E_0$($S^z$)~\footnote{Because the selected magnetic field does not necessarily ensure $E_0$($S^z$$+$1) = $E_0$($S^z$$-$1). }, where $E_0$($S^z$) presents the GS energy in the $S^z$ subspace. Above $H_{\mathrm{c}2}$, $\Delta E$($S^z$) strongly depends on the cluster length along the $b$ axis $n_b$ [see Table ~\ref{tab2}],  thus suggesting the formation of long-range entanglements along the chain B [the Cu$^{2+}$ spins of chain A are nearly fully polarized at $H$ $>$ $H_{\mathrm{c}2}$, see Figs.~\ref{fig3}(c) and \ref{fig3}(d)]. As shown in Figs.~\ref{figr1}(c) and \ref{figr1}(d), the calculated spin gaps of both plateaus, $\Delta E$($n$/3) and $\Delta E$(5$n$/12), tend to zero in the thermodynamic limit, and thus these plateaus may be not stable in the isotropic exchange model no.1. On the other hand, the magnetic-field-induced transitions at $H_{\mathrm{c}2}$ and $H_{\mathrm{c}3}$ are apparent in the experimental magnetization at 1.5 K (Fig.~\ref{fig6}). Thereby, it is still an open question whether additional interaction terms (e.g. DM exchange) are necessary to understand the experimental observations of these magnetic-field-induced transitions at low temperatures.

In sharp contrast, the classical calculations (i.e. CMC) with the same Heisenberg exchange parameters of model on.1 completely fail to reproduce these putative fractional magnetization plateaus and  magnetic-field-induced transitions (Fig.~\ref{fig2}). This suggests that the quantum fluctuations play a critical role in stabilizing these novel plateau phases and the magnetic transitions among them in Cu$_2$(OH)$_3$NO$_3$.

\subsection{Short-range resonating-valence-bond correlations}

In Cu$_2$(OH)$_3$NO$_3$, the possibility of the conventional long-range antiferromagnetic order with a strong order parameter is precluded by the absence of visible magnetic reflections in neutron diffraction~\cite{drillon1995recent} [see Fig.~\ref{figs1}(b)] and the relatively broad/weak anomalies seen in the specific heat and susceptibilities (see Fig.~\ref{fig4}). No clear remanent magnetization at 0 T is observed (see Fig.~\ref{figr4}), suggesting the absence of the conventional ferromagnetic order. Furthermore, the spin-glass freezing is also excluded, as the ac susceptibilities show no obvious frequency dependence [see Fig.~\ref{fig4}(c)]. All of these observations consistently point to a very unconventional magnetic GS in Cu$_2$(OH)$_3$NO$_3$ at 0 T.

Figure~\ref{fig3} shows the local correlation functions $\langle\bm{\mathrm{s}}_i$$\cdot$$\bm{\mathrm{s}}_j\rangle$ of the NN spin pairs $\langle ij\rangle$ calculated by model no.1, and no evident NN dimer freezing is observed,  $\langle\bm{\mathrm{s}}_i$$\cdot$$\bm{\mathrm{s}}_j\rangle$ $\ge$ $-$0.42, which is essentially different from the cases in the $s$ = 1/2 frustrated systems with quenched nonmagnetic impurities~\cite{PhysRevB.68.224416} or bond randomness~\cite{PhysRevB.105.024418} (where the strongest correlation is $\langle\bm{\mathrm{s}}_i$$\cdot$$\bm{\mathrm{s}}_j\rangle$ $\sim$ $-$0.7 $\to$ $-$3/4). Therefore, neither the static VB crystal nor VB glass theory~\cite{PhysRevLett.104.177203} is suitable for understanding the GS nature of the spin system of Cu$_2$(OH)$_3$NO$_3$. The applied magnetic field will further increase $\langle\bm{\mathrm{s}}_i$$\cdot$$\bm{\mathrm{s}}_j\rangle$, and gradually modify the dimension of the spin system from quasi 2 to quasi 1 (see Fig.~\ref{fig3}).

As a result, it may be practical to represent the zero-field GS wavefunction $\mid$GS$\rangle$ in the VB basis. Both the low-$T$ magnetization measured below 5 T and calculations of model no.1 show a quasi-spin-gapped nature (see above), and the LD/ED simulations have marginal finite-size effect at 0 T, thus implying the dominant short-range spin-spin correlations. Thereby, we further restrict ourselves to the NN VB basis, similar to Ref.~\cite{PhysRevB.68.224416}.  We represent the zero-field GS wavefunctions calculated on the  1$\times$3$\times$4,  2$\times$2$\times$4, and 2$\times$3$\times$4 clusters with PBC in the corresponding NN VB bases, and the results are highly similar. For the largest cluster, there exist 60744
different NN VB  states $\mid$VB$_k$$\rangle$ = $2^{-n/4}\prod_{i=1}^{n/2}$($\mid\uparrow_i\downarrow_{j_i}\rangle$$-$$\mid\downarrow_i\uparrow_{j_i}\rangle$), where each spin $i$ forms a singlet/VB with one of its nearest neighbors $j_i$ [see Fig.~\ref{fig7}(a)]. However, these VB  states are nonorthogonal~\cite{PhysRevLett.61.365}, and construct a $N_\mathrm{VB}$ = 52104-dimensional Hilbert space with the orthonormal basis $\mid$VB$'_{k'}$$\rangle$ = $\sum_k$$C_{k'k}$$\mid$VB$_k$$\rangle$. We approximately represent $\mid$GS$\rangle$ in the NN VB basis by
\begin{equation}
|\mathrm{GS}\rangle\approx\sum_{k',k}\langle\mathrm{VB}'_{k'}|\mathrm{GS}\rangle C_{k'k}|\mathrm{VB}_k\rangle.
\label{eq2}
\end{equation}
Here, $\sum_{k'}|\langle\mathrm{VB}'_{k'}$$\mid$$\mathrm{GS}\rangle|^2$ = 0.882, 0.995 for the 1$\times$3$\times$4, 2$\times$2$\times$4 clusters~\footnote{The completeness can be exactly calculated only for the smaller clusters due to the total memory limit $\sim$ 0.8 TB, e.g., 1$\times$3$\times$4 and 2$\times$2$\times$4, where $N_\mathrm{VB}$ = 36 and 1309 are small.}, respectively, suggesting the representation of $\mid$GS$\rangle$ in the NN VB basis is nearly complete.

In the triangular-lattice magnet of Cu$_2$(OH)$_3$NO$_3$, the exchange interactions are spatially anisotropic [see Fig.~\ref{fig1}(c) and Table~\ref{tab1}]. The GS spin-spin correlations calculated by the model no.1 are of course anisotropic as well, but nearly invariant under the space group symmetry due to the tiny $M_0$ [see Fig.~\ref{fig3}(a)]. As a result, $\mid$GS$\rangle$ of model no.1 is not an exactly equal-amplitude superposition of all $\mid$VB$_k$$\rangle$, which is essentially different from short-range spin liquids in isotropic cases, and is a prominent feature of the spatially anisotropic counterparts. More specifically, the NN VB states with VBs along the $b$ axis have larger amplitudes $|C'_k|$ = $|\sum_{k'}\langle\mathrm{VB}'_{k'}$$\mid$$\mathrm{GS}\rangle C_{k'k}|$ [see Fig.~\ref{fig7}(a) for example], at $\mid$GS$\rangle$ of model no.1.  The completeness of the representation of the GS wavefunction in the RVB basis indicates the existence of dominant NN RVB correlations, and thus the low-$T$ phase of the spin system of Cu$_2$(OH)$_3$NO$_3$ may be in the vicinity of a short-range QSL.

\begin{figure}
\begin{center}
\includegraphics[width=8cm,angle=0]{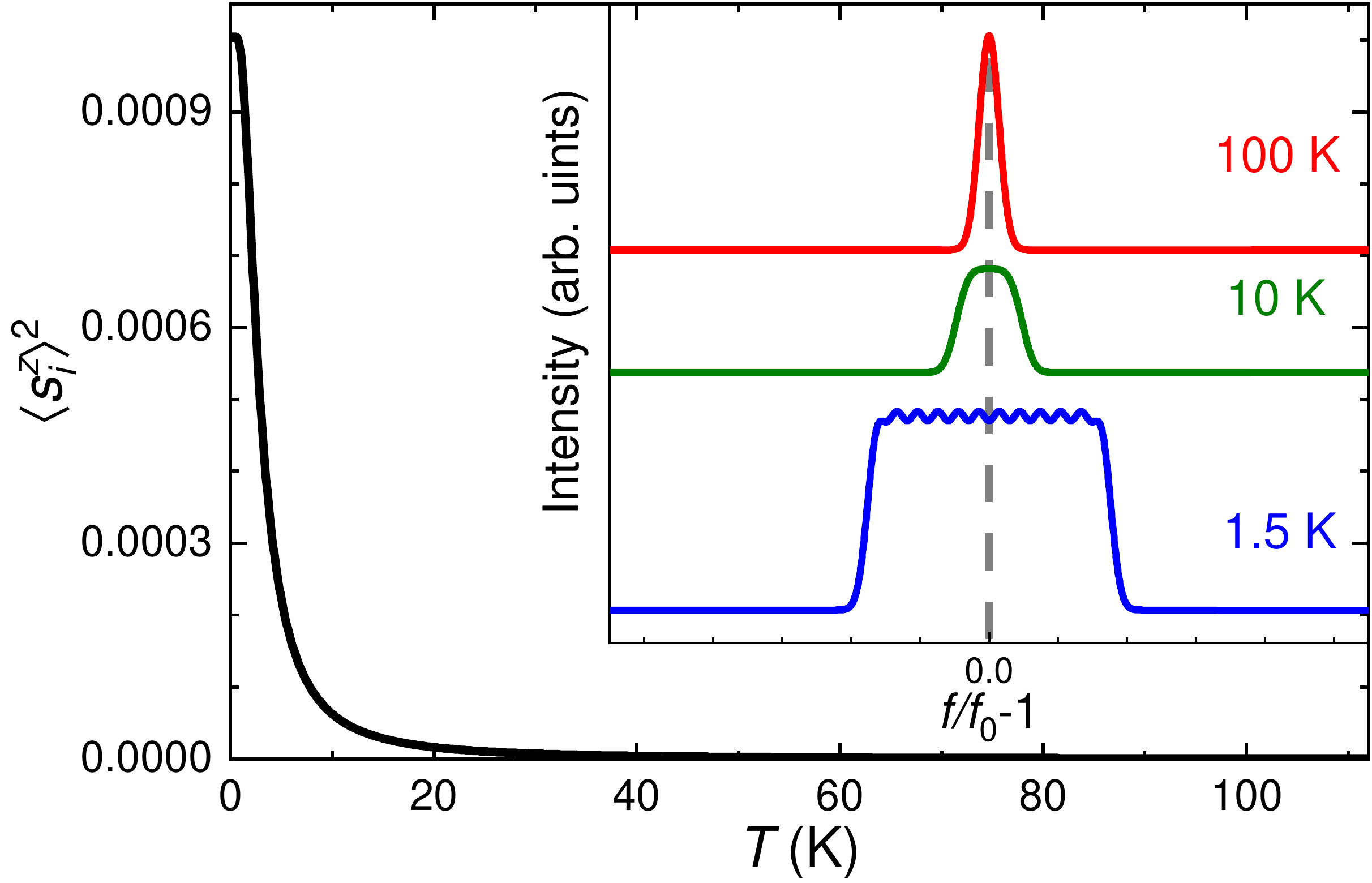}
\caption{The square of the zigzag stripe ordered moment calculated by using the model no.1. The inset shows the simulations of H-NMR spectra. There exist three different Wyckoff positions of H (H1, H2, and H3) and two different Cu positions (Cu1 and Cu2) in Cu$_2$(OH)$_3$NO$_3$, and thus the hyperfine coupling between the H nuclear and local Cu$^{2+}$ electronic spins is distributed, which may account for the broadening of the spectrum below $T_\mathrm{N}$~\cite{kikuchi2018magnetic}, instead of the splitting. $f_0$ = $^{1}\gamma_\mathrm{n}\mu_0H$ and $f$ represent the reference and measuring frequencies, respectively.}
\label{figr6}
\end{center}
\end{figure}

To simulate the wave-vector ($\mathbf{Q}$) dependence of neutron scattering intensity, we further calculate the structure factor of the ground state by
\begin{equation}
S^{zz}(\mathbf{Q})=\frac{1}{n^2}\sum_{j,j'}\langle\mathrm{GS}|s_j^zs_{j'}^z|\mathrm{GS}\rangle\exp[i\mathbf{Q}\cdot(\mathbf{R}_j-\mathbf{R}_{j'})],
\label{eqr1}
\end{equation}
where $\mathbf{R}_j$ is the position vector of the $j^{th}$ spin. As shown in Fig.~\ref{fig7}, the continuum of excitations broadly distributed in $\mathbf{Q}$ space exhibits relatively weak finite-size effects. The $\mathbf{Q}$ dependence along [0, K, 0] is more pronounced than that along [H, 0, 0], suggesting the short-range antiferromagnetic correlations are mainly along the $b$ axis because of the strongest antiferromagnetic $J_{1\mathrm{b}}$. However,  at present the smoking gun experimental evidence of the RVB correlations remains absent, which requires further investigations on Cu$_2$(OH)$_3$NO$_3$.

\subsection{Discussion}

\begin{figure}
\begin{center}
\includegraphics[width=8.3cm,angle=0]{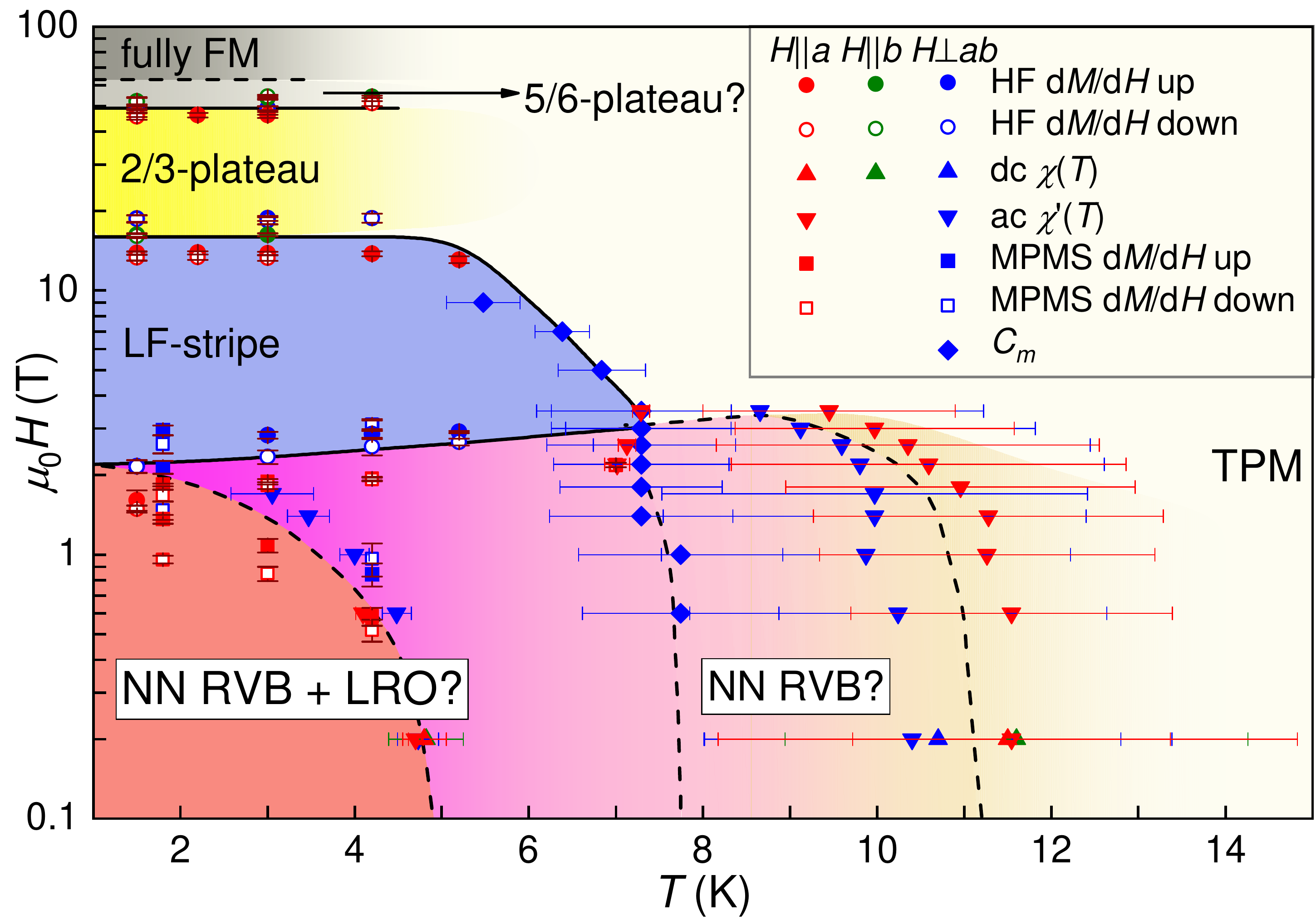}
\caption{$T$-$H$ phase diagram of Cu$_2$(OH)$_3$NO$_3$. The high-field (HF) and MPMS magnetization ($M$) with the applied field ($H$) sweeping up and down, dc and ac susceptibilities ($\chi$ and $\chi'$), and specific heat ($C_m$) are measured on the high-quality single-crystal sample. Error bar on each scatter presents the width of the peak/anomaly observed in the $H$ or $T$  dependence of d$M$/d$H$ or $C_m$ (i.e. the second derivative of the free energy). The spin system is at the thermally induced paramagnetic state (TPM) at high temperatures, and expected to form both dominant NN RVB correlations and weak long-range zigzag-stripe order (LRO) at $\sim$ 0 T and low temperatures, according to the model no.1.}
\label{fig8}
\end{center}
\end{figure}

\begin{figure}
\begin{center}
\includegraphics[width=7cm,angle=0]{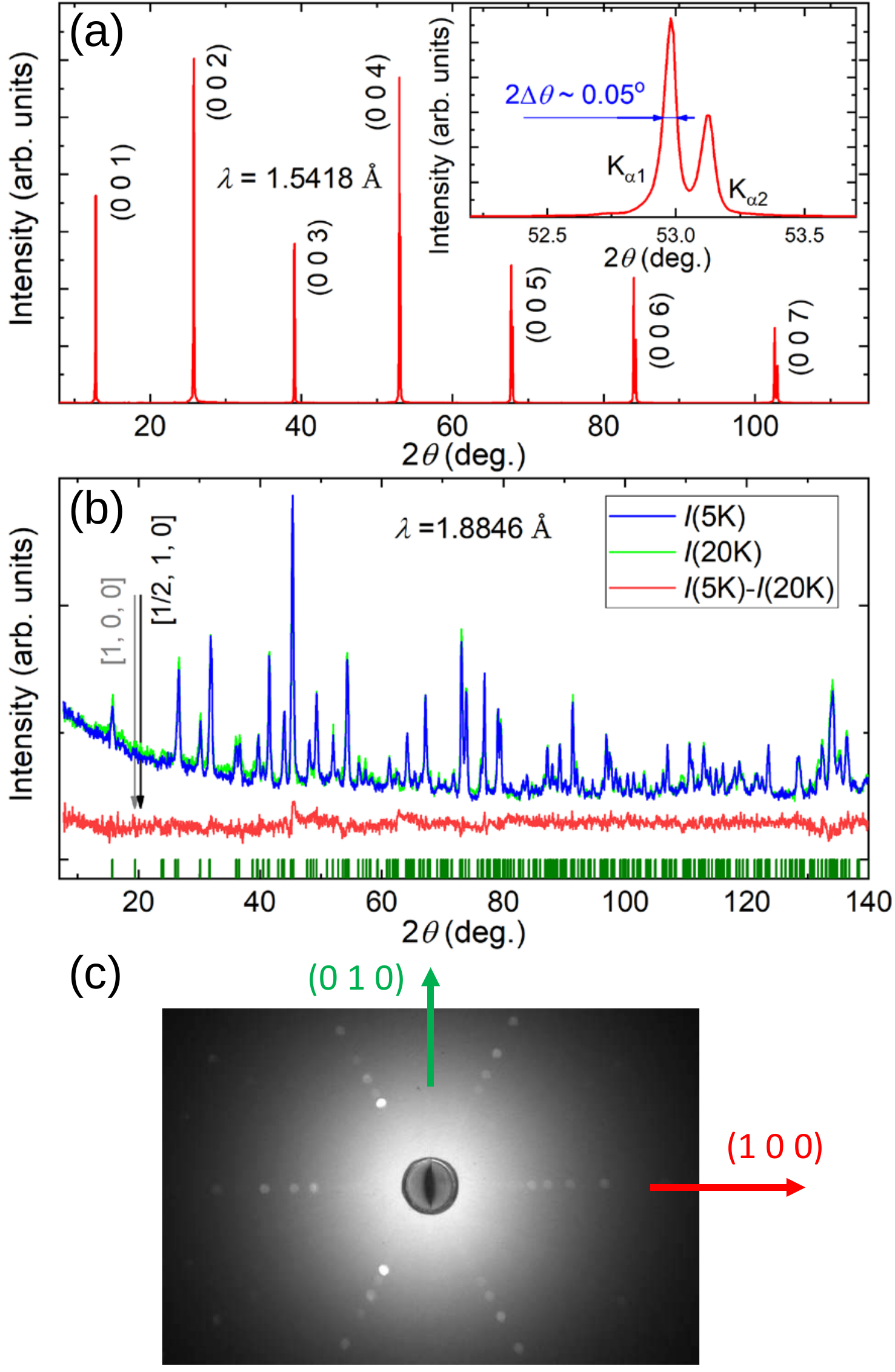}
\caption{(a) Single-crystal XRD ($\lambda$ = 1.5418 \AA) measured on the $ab$-plane of Cu$_2$(OH)$_3$NO$_3$. The inset shows the full widths of half maximum (FWHM) of the (0 0 4) reflection, 2$\Delta$$\theta$ $\sim$ 0.05$^{\circ}$. (b) Neutron diffraction patterns measured above and below $T_\mathrm{N}$ = 8 K, on the powder sample ground from the single crystals of Cu$_2$(OH)$_3$NO$_3$. The red line shows the difference, the arrows mark the theoretically expected magnetic reflections at \textbf{Q} = [1/2, 1, 0] and [1, 0, 0], respectively, in the triangular plane, and the bars display the structural reflections. (c) The Laue XRD pattern.}
\label{figs1}
\end{center}
\end{figure}

Quantum simulations of the refined Heisenberg model no.1 show semi-quantitative agreement with the experimental observations on the high-quality single crystal of Cu$_2$(OH)$_3$NO$_3$. (1) The thermodynamic data, including the magnetic field dependence of magnetization measured at 4.2 and 1.5 K up to 55 T and the temperature dependence of magnetic susceptibilities and specific heat, are semi-quantitatively described by the same model no.1 (see Fig.~\ref{fig2}). (2)  The GS wavefunction given by the same model suggests a long-range zigzag-stripe antiferromagnetic order with a weak collinear sublattice magnetization $M_0$ $\sim$ 0.02 [see Fig.~\ref{fig3}(a)], which may explain the antiferromagnetic order seen in specific heat, susceptibilities  (see Fig.~\ref{fig4}), as well as possibly in the reported NMR spin-lattice relaxation rates on the powder sample~\cite{kikuchi2018magnetic}. (3) The calculated $M_0$ $\sim$ 0.02 is smaller than the sensitivity of the neutron diffraction method and thus qualitatively accounts for the absence of visible magnetic reflections reported in previous~\cite{drillon1995recent} and present works [see Fig.~\ref{figs1}(b)]. (4) The low-field magnetization shows a spin-flop transition at $H_{\mathrm{c}1}$ applied along the $a$ axis and perpendicular to the $ab$ plane, semi-quantitatively consistent with the simulation of model no.1 (see Fig.~\ref{fig5}). (5) Two high-field induced phase transitions at $H_{\mathrm{c}2}$ and $H_{\mathrm{c}3}$ are clearly observed in the magnetization measured at low $T$ up to 55 T (see Fig.~\ref{fig6}), semi-quantitatively consistent with the calculations of model no.1 with the cluster size up to $n$ = 36 (see Table~\ref{tab2}), thus suggesting the possible existence of 2/3- and 5/6-plateau states stabilized by quantum spin fluctuations. (6) The model no.1 suggests the onset of the staggered magnetic moments at low temperatures, and can qualitatively account for the apparent broadening of the NMR spectrum reported in Ref.~\cite{kikuchi2018magnetic} (Fig.~\ref{figr6}).

Given that the minimum model no.1 provides an apt description of the quantum magnetism of Cu$_2$(OH)$_3$NO$_3$, we are able to experimentally determine the temperature-field phase diagram (see Fig.~\ref{fig8}), with the benefit of the many-body computation. The first-order nature of the field-induced magnetic transition at $H_{\mathrm{c}1}$ is evidenced by the asymmetric magnetic hysteresis. In contrast, the field-induced magnetic transitions at $H_{\mathrm{c}2}$ and $H_{\mathrm{c}3}$ are second-order, as no significant difference of $H_{\mathrm{c}2}$ (or $H_{\mathrm{c}3}$) is found between up and down sweeps of the pulsed field (see Fig.~\ref{fig8}).

Despite the success of model no.1, our thermodynamic quantities measured on the high-quality single crystal of Cu$_2$(OH)$_3$NO$_3$ clearly exhibit the anisotropic nature in the spin space at low temperatures (see Fig.~\ref{fig8}), which is, of course, beyond any Heisenberg models. Besides that, the measured magnetic susceptibilities and specific heat show deviations from the Heisenberg model no.1 below $\sim$ 10 K [Figs.~\ref{fig2}(b) and \ref{fig2}(d)]. These observations indicate the existence of higher-order anisotropic superexchange of $\sim$ 0.1$J_1$ (see Appendix~\ref{sec3}) due to the weak spin-orbit coupling of the 3$d$ electron~\cite{PhysRev.120.91}, which also requires further investigation on Cu$_2$(OH)$_3$NO$_3$.

There exist six different NN interaction parameters allowed by the $P$2$_1$ space group symmetry of Cu$_2$(OH)$_3$NO$_3$, and thus it seems unrealistic to calculate the complete ground-state phase diagram in the complex interaction parameter space. One might simplify the problem, for instance, by assuming $J_{1\mathrm{a}}$ $\sim$ $J_{1\mathrm{b}}$ $\sim$ $J$ and $J_{1\mathrm{c}}$ $\sim$ $J_{1\mathrm{d}}$ $\sim$ $J_{1\mathrm{e}}$ $\sim$ $J_{1\mathrm{f}}$ $\sim$ $J'$, and obtain the well-known $J$-$J'$ model on the spatially anisotropic triangular lattice~\cite{PhysRevB.74.014408}. Spin-liquid phases are expected, as long as the spatial anisotropy is relatively strong, $J'$/$J$ $\lesssim$ 0.8~\cite{PhysRevB.74.014408}. Thereby, a spin-liquid ground state is very likely to form in the spatially anisotropic triangular-lattice model. Loosely speaking, the model no.1 may be not far away from the $J$-$J'$ model, as the strongest antiferromagnetic couplings $J_{1\mathrm{b}}$ $\sim$ 70 K  and $J_{1\mathrm{a}}$ $\sim$ 7 K are indeed along the ``chain'' direction (the $b$ axis), and other ``interchain'' couplings are either much weaker or ferromagnetic. Moreover, the spin gap calculated by the model no.1 is very small in the thermodynamic limit [see Fig.~\ref{figr1}(b)], which also seems to be consistent with the predictions of the $J$-$J'$ model with $J'$/$J$ $\lesssim$ 0.8~\cite{PhysRevB.74.014408}. Essentially, both the model no.1 used in this work and the previous $J$-$J'$ model suggest that the strength of the spatial anisotropy can increase quantum fluctuations and can suppress conventional magnetic order.

\section{Conclusions}

We performed an extensive single-crystal study on the $s$ = 1/2 spatially anisotropic triangular-lattice quantum magnet of Cu$_2$(OH)$_3$NO$_3$ without evident structural disorder. The effect spin Hamiltonian with symmetry-allowed NN Heisenberg exchange interactions is experimentally determined by fitting the magnetization measured up to 55 T, and shows semi-quantitative agreement with the experimental observations, including the $T$ dependence of magnetic susceptibilities and specific heat, the weak long-range antiferromagnetic transition, the absence of visible magnetic reflections in neutron diffraction, the quasi-spin-gapped behavior and three magnetic-field-induced quantum phase transitions seen in the magnetization. Both the dominant short-range resonating-valence-bond correlations and weak long-range stripe order coexist  in the GS wavefunction obtained by diagonalizing the same spin Hamiltonian, thus suggesting the magnetic GS of Cu$_2$(OH)$_3$NO$_3$ may be in the vicinity of a short-range QSL. Our work sheds light on the search for QSLs in structurally disorder-free quantum magnets with spatially anisotropic exchange interactions.

\begin{table}[h]
\caption{The crystal structure refined from the single-crystal XRD data measured on Cu$_2$(OH)$_3$NO$_3$ at 300 K, with the comparison of the DFT result. The space group of $P2_1$, and lattice parameters, $a$ =  5.5964(2) \AA, $b$ =   6.0784(3) \AA, $c$ =   6.9251(3) \AA, and $\beta = 94.552(4)^{\circ}$, measured by XRD are fixed in our DFT simulations. $R$($F$) ($I$ $>$ 3$\sigma$) = 6.4\% and $R_{w}$($F$) ($I$ $>$ 3$\sigma$) = 6.8\% based on 1414 reflections with $h$ range of $-7 \rightarrow 7$, $k$ range of $-8 \rightarrow 8$, and $l$ range of $-9 \rightarrow 9$, are resulted in the structure refinement.}
\begin{center}
\begin{tabular}{l||l|l}
\hline
\hline
 & XRD (300 K) & DFT \\
\hline
Cu1: $x$, $y$, $z$&0.9985(3), 0.93(1), 0.9938(2)&0.999, 0.939, 0.991\\
100$\times U_{\mathrm{iso}}$ & 0.57(6) & - \\
Cu2: $x$, $y$, $z$&0.5093(2), 0.19(1), 0.9978(2)&0.502, 0.190, 0.990\\
100$\times U_{\mathrm{iso}}$ & 0.54(6) & - \\
\hline
N: $x$, $y$, $z$&0.229(2), 0.20(1), 0.405(1)&0.237, 0.181, 0.428 \\
100$\times U_{\mathrm{iso}}$ & 1.2(4) & - \\
\hline
O1: $x$, $y$, $z$&0.874(1), 0.18(1), 0.8565(9)&0.875, 0.194, 0.855\\
100$\times U_{\mathrm{iso}}$ & 0.8(3) & - \\
O2: $x$, $y$, $z$&0.312(1), 0.94(1), 0.8820(9)&0.313, 0.946, 0.880\\
100$\times U_{\mathrm{iso}}$ & 0.6(3) & - \\
O3: $x$, $y$, $z$&0.689(1), 0.92(1), 0.114(1)&0.695, 0.934, 0.114\\
100$\times U_{\mathrm{iso}}$ & 0.9(4) & - \\
O4: $x$, $y$, $z$&0.211(1), 0.19(1), 0.2309(9)&0.214, 0.192, 0.244\\
100$\times U_{\mathrm{iso}}$ & 1.3(4) & - \\
O5: $x$, $y$, $z$&0.382(2), 0.09(1), 0.493(1)&0.373, 0.037, 0.509\\
100$\times U_{\mathrm{iso}}$ & 4.1(7) & - \\
O6: $x$, $y$, $z$&0.902(2), 0.81(1), 0.510(1)&0.880, 0.815, 0.475\\
100$\times U_{\mathrm{iso}}$ & 3.4(6) & - \\
\hline
H1: $x$, $y$, $z$& - &0.925, 0.204, 0.721\\
H2: $x$, $y$, $z$& - &0.312, 0.028, 0.737\\
H3: $x$, $y$, $z$& - &0.728, 0.092, 0.256\\
\hline
\hline
\end{tabular}
\end{center}
\label{tabs1}
\end{table}

\begin{acknowledgments}
We thank Haijun Liao for helpful discussion. This work was supported by the Fundamental Research Funds for the Central Universities, HUST: 2020kfyXJJS054. Benqiong Liu was supported by the National Natural Science Foundation of China (Grant No.11875238).
\end{acknowledgments}

\appendix

\section{Structural characterization of the single crystal of Cu$_2$(OH)$_3$NO$_3$}
\label{sec1}

Both the single-crystal and Laue XRD measured on the $ab$-plane show very narrow (FWHM $\sim$ 0.05$^{\circ}$ in 2$\theta$) reflections (see Fig.~\ref{figs1}), implying the high quality of the single crystal of Cu$_2$(OH)$_3$NO$_3$. From fitting the intensities of the single-crystal reflections in GSAS program~\cite{gsas}, we obtained the crystal structure of Cu$_2$(OH)$_3$NO$_3$ (see Table~\ref{tabs1}).

\section{Strengths of Heisenberg exchange couplings estimated by DFT+$U$}
\label{sec2}

Starting from the XRD result, we first constructed the positions of H atoms, and then optimized the geometry (see Table~\ref{tabs1}) using DFT in VASP with the generalized gradient approximation~\cite{PhysRevLett.77.3865}. The 6$\times$6$\times$6 k-mesh was used, the lattice parameters were fixed to the experimental values, and the residual forces were below 0.006 eV/\AA~in the fully optimized structures.

The exchange integrals monotonically decrease with the increase of Coulomb repulsion $U_\mathrm{eff}$. For Cu$_2$(OH)$_3$NO$_3$, $U_\mathrm{eff}$ was fixed to its typical value of 10.5 eV~\cite{PhysRevLett.117.037206} in our DFT+$U$ calculations. We considered the supercell of 1$\times$1$\times$2 unit cells, and calculated the total energies of the interlayer ferromagnetic and antiferromagnetic states, $E$($\mid\uparrow\uparrow\uparrow\uparrow\uparrow\uparrow\uparrow\uparrow\rangle$) = $-$236.532784 eV and $E$($\mid\uparrow\uparrow\uparrow\uparrow\downarrow\downarrow\downarrow\downarrow\rangle$) = $-$236.532760 eV, respectively. The strength of interlayer coupling was obtained by $J_{\perp}$ = [$E$($\mid\uparrow\uparrow\uparrow\uparrow\uparrow\uparrow\uparrow\uparrow\rangle$)$-$$E$($\mid\uparrow\uparrow\uparrow\uparrow\downarrow\downarrow\downarrow\downarrow\rangle$)]/4 = $-$0.07 K (ferromagnetic), in agreement with the previously reported DFT result~\cite{ruiz2006theoretical}, and suggesting the quasi two-dimensional nature of the spin system of Cu$_2$(OH)$_3$NO$_3$. Similarly, we obtained the strengths of intra-layer NN and second-neighbor exchange couplings by considering another supercell of 2$\times$1$\times$1 unit cells, as summarized in Table~\ref{tabs2}.

\begin{figure}
\begin{center}
\includegraphics[width=7.5cm,angle=0]{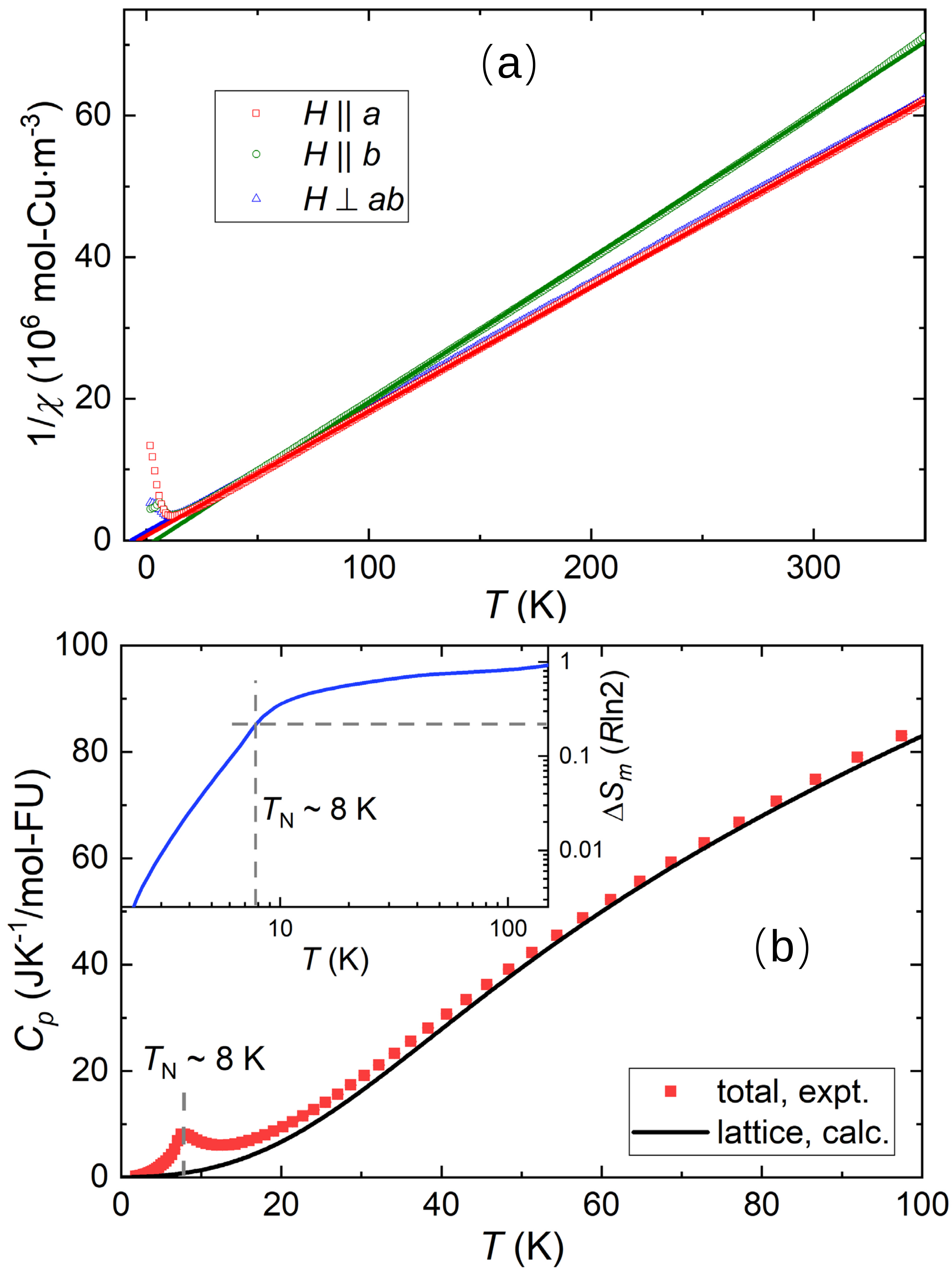}
\caption{(a) Inverse magnetic susceptibilities measured in the field of 0.2 T parallel to the $a$ and $b$ axes and perpendicular to the $ab$ plane, respectively. The colored lines display the Curie-Weiss fits from 50 to 350 K. (b) Zero-field specific heat measured on the single crystal of Cu$_2$(OH)$_3$NO$_3$. The black line shows the lattice contribution calculated by VASP+Phonopy. The inset presents the magnetic entropy increase from 1.8 K (per mole Cu$^{2+}$), where the dashed lines show $\Delta S_m$ $\sim$ 22\%$R$ln2 at $T_\mathrm{N}$ $\sim$ 8 K.}
\label{figs2}
\end{center}
\end{figure}

\begin{table}
\caption{Cuire-Weiss fittings of the $T$ dependence of the magnetic susceptibilities, $\chi$ = $C_\mathrm{CW}$/($T$$-$$\theta_\mathrm{CW}$).}
\begin{center}
\begin{tabular}{ l  ||  l  |  l  |  l }
\hline
\hline
 & $H$ $||$ $a$ & $H$ $||$ $b$ & $H$ $\perp$ $ab$ \\
 \hline
 \multicolumn{4}{c}{Temperature range of the fits: 20-350 K}\\
\hline
$C_\mathrm{CW}$ (10$^{-6}$ m$^3$K$^{-1}$/mol-Cu) & 5.681(2) & 5.043(2) & 5.688(3) \\
$g$ = 2$\sqrt{\frac{k_\mathrm{B}C_\mathrm{CW}}{\mu_0N_\mathrm{A}\mu_\mathrm{B}^2}}$ & 2.1956(4) & 2.0685(4) & 2.1968(7) \\
$\theta_\mathrm{CW}$ (K) & $-$3.38(8) & 0.67(6) & $-$6.2(1) \\
Adj. R-Square~\footnote{see https://www.originlab.com/doc/Origin-Help/Interpret-Regression-Result for Adj. R-Square.} & 0.99996 & 0.99997 & 0.99988 \\
\hline
 \multicolumn{4}{c}{Temperature range of the fits: 50-350 K}\\
\hline
$C_\mathrm{CW}$ (10$^{-6}$ m$^3$K$^{-1}$/mol-Cu) & 5.690(2) & 4.917(4) & 5.705(4) \\
$g$ = 2$\sqrt{\frac{k_\mathrm{B}C_\mathrm{CW}}{\mu_0N_\mathrm{A}\mu_\mathrm{B}^2}}$ & 2.1972(4) & 2.0425(8) & 2.2001(7) \\
$\theta_\mathrm{CW}$ (K) & $-$3.76(9) & 3.9(2) & $-$6.9(1) \\
Adj. R-Square & 0.99995 & 0.99982 & 0.99988 \\
\hline
\hline
\end{tabular}
\end{center}
\label{tabs3}
\end{table}

\begin{table*}
\caption{Collinear magnetic structures and their energy differences from that of the fully ferromagnetic state (supercell: 2$\times$1$\times$1 unit cells). From fitting the energy differences, we estimate both NN ($J_{1\mathrm{a}}$-$J_{1\mathrm{f}}$) and second-neighbor ($J_2$) exchange couplings. The least-squares standard deviation is defined by $\sigma$ = $\sqrt{\frac{1}{N_{\mathrm{st}}}\sum_i(X_i^{\mathrm{DFT}}-X_i^{\mathrm{fit}})^2}$, where $X_i^{\mathrm{DFT}}$ and $X_i^{\mathrm{fit}}$ are the DFT and fitted values, respectively, $N_{\mathrm{st}}$ = 11 is the number of the non-fully polarized states taken into account.}
\begin{center}
\begin{tabular}{ l || l | l | l | l | l | l | l | l | l | l | l}
    \hline
    \hline
st = $\mid S_1^zS_2^zS_3^zS_4^z$ &
$\mid\downarrow\uparrow\uparrow\uparrow$ &
$\mid\uparrow\uparrow\downarrow\uparrow$ &
$\mid\downarrow\downarrow\uparrow\uparrow$ &
$\mid\downarrow\uparrow\downarrow\uparrow$ &
$\mid\downarrow\downarrow\downarrow\uparrow$ &
$\mid\downarrow\downarrow\downarrow\downarrow$ &
$\mid\uparrow\downarrow\uparrow\downarrow$ &
$\mid\downarrow\uparrow\uparrow\downarrow$ &
$\mid\downarrow\uparrow\uparrow\downarrow$ &
$\mid\downarrow\uparrow\uparrow\uparrow$ &
$\mid\downarrow\uparrow\uparrow\uparrow$\\
\quad\quad\quad$S_5^zS_6^zS_7^zS_8^z\rangle$ &
\quad$\uparrow\uparrow\uparrow\uparrow\rangle$ &
\quad$\uparrow\uparrow\uparrow\uparrow\rangle$ &
\quad$\uparrow\uparrow\uparrow\uparrow\rangle$ &
\quad$\uparrow\uparrow\uparrow\uparrow\rangle$ &
\quad$\uparrow\uparrow\uparrow\uparrow\rangle$ &
\quad$\uparrow\uparrow\uparrow\uparrow\rangle$ &
\quad$\uparrow\downarrow\uparrow\downarrow\rangle$ &
\quad$\uparrow\downarrow\uparrow\uparrow\rangle$ &
\quad$\uparrow\uparrow\uparrow\downarrow\rangle$ &
\quad$\downarrow\uparrow\uparrow\uparrow\rangle$ &
\quad$\uparrow\uparrow\downarrow\uparrow\rangle$\\
    \hline
DFT + $U$, & 12.3 & 71.9 & 19.4 & 75.5 & 68.5 & 17.7 & 127.9 & 93.9 & 131.8 & 25.9 & 71.8\\
$E$($\mid\uparrow\uparrow\uparrow\uparrow\uparrow\uparrow\uparrow\uparrow\rangle$)$-E$(st) (K) &&&&&&&&&&&\\
    \hline
least-squares fitted, & 12.6 & 65.2 & 19.4 & 74.4 & 68.1 & 20.7 & 130.2 & 93.4 & 134.9 & 23.9 & 69.8\\
$E$($\mid\uparrow\uparrow\uparrow\uparrow\uparrow\uparrow\uparrow\uparrow\rangle$)$-E$(st) (K) &&&&&&&&&&&\\
    \hline
exchange couplings & \multicolumn{11}{c}{$J_{1\mathrm{a}}$ = 3 K, $J_{1\mathrm{b}}$ = 60 K, $J_{1\mathrm{c}}$ = 7 K, $J_{1\mathrm{d}}$ = $-$7 K, $J_{1\mathrm{e}}$ = 3 K, $J_{1\mathrm{f}}$ = 10 K, $J_{2}$ = 0.7 K and $\sigma$ = 2.7 K} \\
and standard deviation & \multicolumn{11}{c}{}\\
    \hline
    \hline
\end{tabular}
\end{center}
\label{tabs2}
\end{table*}

\section{$g$ factors and Weiss temperature of Cu$_2$(OH)$_3$NO$_3$}
\label{sec3}

From the Curie-Weiss fittings [Fig.~\ref{figs2}(a)], we got the Cuire constants ($C_\mathrm{CW}$) and Weiss temperatures ($\theta_\mathrm{CW}$) parallel to the $a$ and $b$ axes, and perpendicular to the $ab$ plane, as listed in Table~\ref{tabs3}. Using $C_\mathrm{CW}$ = $\frac{\mu_0N_\mathrm{A}\mu_\mathrm{B}^2g^2}{4k_\mathrm{B}}$, we further obtained the $g$ factors, $g_a$ $\sim$ 2.20,  $g_b$ $\sim$ 2.06, $g_\perp$ $\sim$ 2.20, and thus the average $g$ = ($g_a$+$g_b$+$g_\perp$)/3 $\sim$ 2.16 used in the main text. The strength of DM interaction was estimated to be $\sim$ $J_1$($g$$-$$g_e$)/$g_e$ $\sim$ 0.1$J_1$ by considering the weak spin-orbit coupling~\cite{PhysRev.120.91}, where $g_e$ = 2.0023 is the free electron $g$ factor. Our susceptibilities ($\chi$) are consistent with those previously reported in Ref.~\cite{linder1995single}, by converting $\chi$ from SI to the Gaussian system.

From fitting, we find that the Weiss temperature is relatively small, $|\theta_\mathrm{CW}|$ $<$ 10 K ($\ll$ $J_{1\mathrm{b}}$ $\sim$ 70 K),  in Cu$_2$(OH)$_3$NO$_3$. This can be attributed to the competing exchange interaction, including both antiferromagnetic and ferromagnetic contributions ($J_{1\mathrm{a}}$, $J_{1\mathrm{b}}$, $J_{1\mathrm{c}}$, $J_{1\mathrm{d}}$, $J_{1\mathrm{e}}$, $J_{1\mathrm{f}}$). The Weiss temperature is roughly consistent with both the model no.1 and the model  no.2 reported in Ref.~\cite{ruiz2006theoretical} in the mean-field approximation, $\theta_\mathrm{CW}$ $\sim$ $-$($J_{1\mathrm{a}}$+$J_{1\mathrm{b}}$+$J_{1\mathrm{c}}$+$J_{1\mathrm{d}}$+$J_{1\mathrm{e}}$+$J_{1\mathrm{f}}$)/4 $\sim$ 7 K (no.1) and $-$2 K (no.2), respectively.  It is worth to mention that in the previously reported work~\cite{ruiz2006theoretical} Ruiz \textit{et al.} refine the couplings by using only the temperature dependence of the magnetic susceptibility measured on powder samples, and thus the resulted Weiss temperature is quite in line with the experimental value. In comparison, we determine the exchange couplings by considering both magnetization and susceptibilities measured on single-crystal samples up to 55 T [see Figs.~\ref{fig2}(a) and \ref{fig2}(b)].

\section{Magnetic specific heat of Cu$_2$(OH)$_3$NO$_3$}
\label{sec4}

The lattice specific heat of Cu$_2$(OH)$_3$NO$_3$ was calculated by VASP+Phonopy, on a supercell of 2$\times$2$\times$1 unit cells (total 96 atoms) [see Fig.~\ref{figs2}(b)]. At high temperatures, the magnetic heat capacity is negligible, and the measured heat capacity is well consistent with the independent $ab$-$initio$ simulation of lattice heat capacity [see Fig.~\ref{figs2}(b)] thus supporting the validity of our experimental data. By subtracting this lattice contribution from the measured total specific heat, we obtained the magnetic specific heat $C_m$ (see main text). The magnetic entropy $T$-integrated from $C_m$/$T$ gets negligible, $\Delta S_m\le$ 0.3\%$R$ln2 [inset of Fig.~\ref{figs2}(b)], indicating that we are indeed accessing the GS property of Cu$_2$(OH)$_3$NO$_3$, at $\sim$ 2 K.

\bigbreak

\bibliography{Cu2OH3NO3}

\end{document}